\documentclass{jfm}
\usepackage{graphicx}
\usepackage{epstopdf, epsfig}

\usepackage{color}
\usepackage{amsmath}
\usepackage{floatrow}
\usepackage[utf8]{inputenc}
\usepackage{hyphenat}

\usepackage{hyperref}
\hypersetup{
    colorlinks=true,
    linkcolor=blue,
    citecolor=blue
    }
\usepackage{float}
\usepackage{csquotes}
\usepackage[utf8]{inputenc}
\usepackage[english]{babel}
\usepackage{natbib}
\usepackage{xcolor}
\usepackage{mathabx}
\usepackage{fdsymbol}
\usepackage{tikz}

\shorttitle{Coherent structures in a contraction} 
\title{Coherent turbulent structures in a rapid contraction}
\author[Alhareth, Mugundhan, Langley \& Thoroddsen]
{Abdullah A. Alhareth\aff{1,2}, 
        Vivek Mugundhan\aff{1,3}, Kenneth R. Langley\aff{1}
        \and \, Sigur$\eth$ur T. Thoroddsen\aff{1}\footnotemark[2]
 \footnotetext[1]{Corresponding author: Sigurdur.Thoroddsen@KAUST.edu.sa} 
}
\affiliation{
\aff{1}Division of Physical Sciences and Engineering,
King Abdullah University of Science and Technology (KAUST),
Thuwal, 23955-6900, Saudi Arabia
\aff{2}Center of Excellence for Aeronautics and Astronautics (CEAA),
King Abdulaziz City for Science and Technology (KACST),
Riyadh 12354, Saudi Arabia
\aff{3}Department of Mechanical Engineering, Amrita School of Engineering Coimbatore, Amrita Vishwa Vidyapeetham, India.}

\begin{document}

\maketitle

\begin{abstract}
The coherent vortical structures in turbulent flow through a strong 16:1 3-D contraction, are studied using time-resolved volumetric measurements.
Visualization using vorticity magnitude criterion shows the emergence of long, stretched cylindrical vortices aligned with the mean flow. 
This alignment is quantified by PDFs of the direction cosines.  We propose two measures to quantify the alignment, the peak height in the probability and a coefficient from the moment of the PDF, both of which reaffirm the strong streamwise alignment.
The r.m.s. streamwise vorticty grows within the contraction to becoming 4.5 times larger than the transverse component, at the downstream location where the contraction ratio $C=11$.  The characteristic vortices become as long as the measurement volume, or more than 4 times the integral scale at the entrance to the contraction.  
We also characterize the vorticity enhancement along individual vortices, measuring 65\% strengthening over the distance where $C$ goes from 4 to 11. 
The prevalence of these coherent structures is estimated from 700,000 measured volumes, showing that near the outlet it is more likely to have 1 or 2 of these structures present than none.
\end{abstract}

\section{Introduction} 

Ever since the seminal work of \cite{BrownRoshko1974} 
the study of turbulent flows has increasingly focused on the presence and role of coherent vortical structures.  Besides the rollers in the mixing layer, other examples are the hairpin structures in the boundary layers \citep{Bandyopadhyay981,Zhou_Adrian_1999,Ganapathisubramani2003,Marusic_2007} as reviewed by \cite{Marusic2019}, vortices in swirling jets \citep{Ianiro2018} and structures in turbulent Taylor-Couette flow \citep{Grossmann2016}. 
\textcolor{black}{In wall-bounded flows, streamwise rolls transport the low-speed fluid near the wall away from it to form long streamwise streaks, which subsequently break down by {\it bursting} (\cite{Kim_Kline_Reynolds_1971}). 
The roll-streak interaction maintains turbulence by the self-sustaining mechanism \citep{Hamilton_Kim_Waleffe_1995, SCHOPPA_HUSSAIN_2002,McKeon_2017}.}
The advent of volumetric measurements of turbulent velocity fields has opened up new avenues for the identification and study of such structures \citep{Elsinga2006,Schroder2023}.  Identifying and characterizing such coherent structures in a variety of turbulent flows should aid in lower-order modeling of their behavior for fundamental elucidation and various applications \citep{Rowley2009,Schmid2010}.

There exist numerous studies of turbulent flow through a contraction, because of its relevance to many industrial flow configurations, as concerns mixing, pressure drop, noise etc.  
\textcolor{black}{\cite{prandtl1933attaining} early on suggested treating random distributions of vorticity as a primary quantity underlying the turbulent velocity fluctuations.  The stretching of strain-aligned vortex filaments amplifies their vorticity.  
By applying continuity and conservation of angular momentum/circulation, in terms of a streamwise contraction ratio $c$, the stretched fluctuating vorticity component in the streamwise direction grows by $c$, while in the compressed transverse directions the fluctuations reduce by a factor of $c^{-1/2}$.
On the other hand, for the corresponding velocity fluctuations, streamwise fluctuations $u_{rms}$ decay as $c^{-1}$ while the transverse $v_{rms}$ and $w_{rms}$ fluctuations grow proportional to $c^{1/2}$.
Similar dependences were obtained from theory by \cite{taylor1935turbulence} and \cite{ribner1952spectrum}.
\cite{Batchelor1954} independently derived such relations assuming rapid distortion of fluid elements relative to their displacement which was later called the {\it Rapid Distortion Theory} (RDT) \citep{Sreenivasan1978,Hunt1990}.  
Experiments, primarily using single-point hot-wire measurements \citep{Uberoi1956, Hussain1976, Tanatichat1980,Ayyalasomayajula2006, Thoroddsen1995} verified qualitatively this behavior for the velocity fluctuations. 
However, for rapid contraction with $c>4$, $u_{rms}$ increased after an initial decay, which \cite{Ertunc2008} have attributed to measurement errors or fluctuating inlet flow.} 


\section{Experimental Setup} 
\subsection{Water tunnel}
\begin{figure} 
\centering
\begin{tabular}{ c c c}
\includegraphics[width=0.35\textwidth]{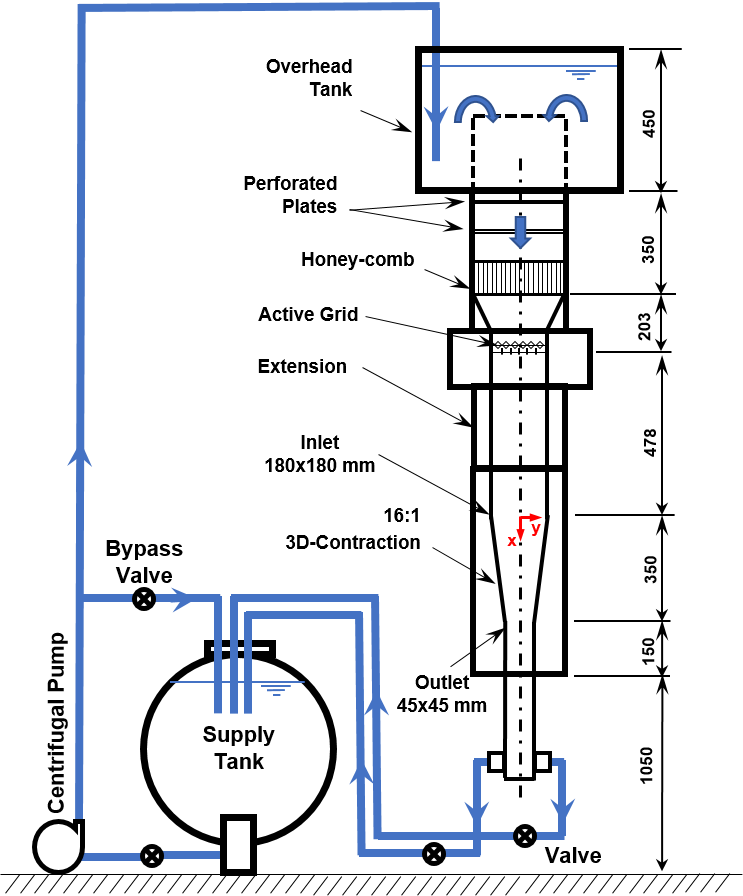} &
\includegraphics[width=0.35\textwidth]{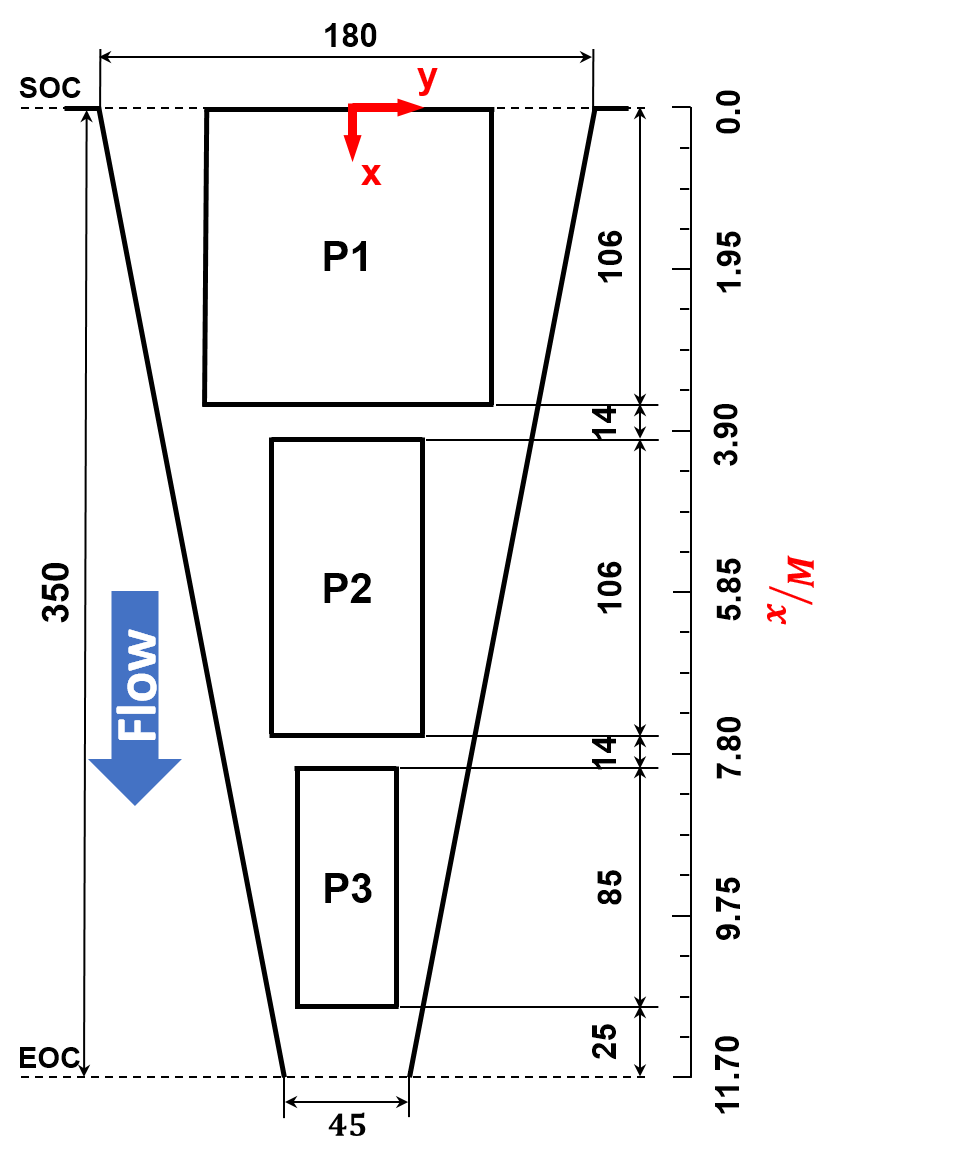}\vspace{-0.1in}&
\includegraphics[width=0.30\textwidth]{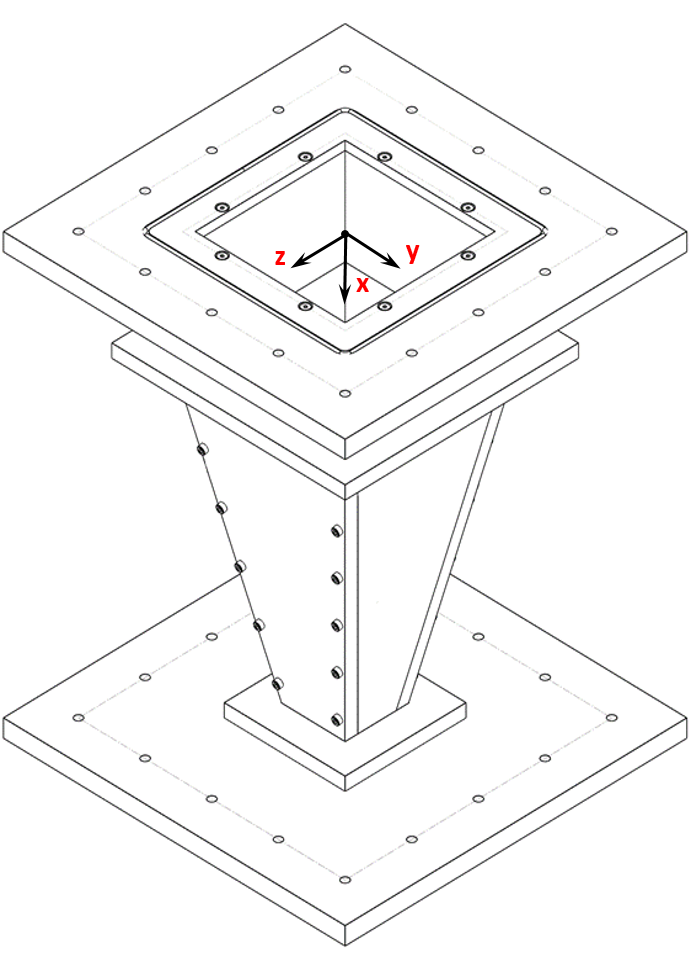} \\
\end{tabular}
\caption {(a) Schematic of the gravity-driven water tunnel with a two-directional 16:1 contraction. 
(b) The three measurement volumes $P1-P3$ with respect to the coordinate axes positioned at the start of the contraction.  The depth of the measurement volumes into the board is 23 mm.  $M$ is the mesh size of the active grid and SOC and EOC represent the start and end of the contraction. (c) Three-dimensional rendering of the contraction. All dimensions in mm.}
\label{Water_Tunnel}
\end{figure}

The water flow facility (Fig. \ref{Water_Tunnel}a) is the same as used by \cite{mugundhan2020alignment}, but with a different contraction section, which now contracts the flow in both horizontal directions.
The 16:1 area contraction reduces the square cross-section from 180$\times$180 mm$^2$ at the inlet to 45$\times$45 mm$^2$ at the outlet.  The water is pumped into a constant-head inlet tank.  Honeycomb and perforated plates kill off large-scale motions and make the flow uniform as it enters the active grid, consisting of 10 independently rotated rods with flat blades (with holes) and a mesh size of $M=30$ mm.  This allows synchronized or random grid rotation protocols to inject turbulent fluctuations, as explained in \cite{mugundhan2020alignment}.  Herein we use only results from the random mode.  Following the active grid we include a 478-mm-long channel of uniform cross-section before the inlet to the contraction, to improve transverse homogeneity.  On the other hand, this long section reduces the turbulence intensity to $u_{rms}/\left\langle U_{\text{in}} \right\rangle = 5.8$\%, while $Re_{\lambda}=192$ at the inlet to the contraction.  The inlet velocity is $\left\langle U_{\text{in}} \right\rangle = 0.28$ m/s, but reaches 2.52 m/s near the bottom of region P3, i.e. closest measurement to the outlet.   Table I lists other turbulent quantities slightly inside the  inlet of the contraction, at $x=5$ mm.  Figure \ref{Water_Tunnel}(b) shows the location of the three measurement volumes, which were investigated in experiments on separate days, as they require rearrangement of the laser-optics and cameras. 


\begin{table}
\centering
\def~{\hphantom{0}}
\centering
\begin{tabular}{c c c c c c c c c c c c}
 \hline

Parameter & & $u_{rms}/{\left \langle U_{in} \right \rangle}$ & ${k}$ & ${\varepsilon}$ & ${L_I}$ & ${\lambda}$ & ${\eta}$ & ${Re_{L}}$ & ${Re_{\lambda}}$ & $\Delta x/\eta$& $S^{*}$ \\
Unit & & $\%$ & $m^2 s^{-2}$ & $m^2 s^{-3}$ & $mm$ & $mm$ & $mm$ & $-$ & $-$ & $-$ & $-$ \\
Value & & $5.77$ & $5.04\times10^{-4}$ & $8.12\times10^{-5}$ & $30.19$ & $8.37$ & $0.33$ & $691$ & $192$ & $2.01$ & $301$ \\

\hline\vspace{-0.35in}\\
\end{tabular}
\caption{Turbulence quantities slightly inside the contraction, at $x=5$ mm.  Mean inlet velocity, ${\left \langle U_{in} \right \rangle}$ $\approx$ 0.28 m/s, streamwise velocity fluctuation, $u_{rms}$, turbulent kinetic energy, ${k}$, dissipation rate, ${\varepsilon}$, streamwise integral length scale, ${L_I}$, and Taylor microscale, ${\lambda}$, computed by two-point spatial correlation function $f(r)=[{\left \langle u(x)u(x+\mathbf{e_x}r)\right \rangle}]/{\left \langle u^2 \right \rangle}$.
Kolmogorov length scale, ${\eta}=(\nu^3/\varepsilon)^{1/4}$. Reynolds numbers based on ${L_I}$ and ${\lambda}$.
The velocity-grid measurement resolution, $\Delta x$. 
$S^{*}$ characterizes the maximum mean strain, in terms of a time-scale ratio from the mean strain and the turbulent straining, as 
$S^{*} = \sqrt{3}\, (\partial \left \langle U \right \rangle/\partial x)\, (k/\varepsilon)$,
with the prefactor for an axisymmetric contraction, according to \cite{Ayyalasomayajula2006}.}
\end{table}
\label{3D-TP}

\subsection{Volumetric velocity measurements}

We employ the 3-D Lagrangian Particle Tracking Velocimetry (LPT) with Shake-the-Box algorithm \citep{schanz2016shake}, as implemented by LaVision (Davis 10.2 software) to obtain time-resolved volumetric flow fields.
We utilize four high-speed 4Mpx video cameras (Phantom V2640) equipped with 85 mm Nikkor tilt lenses set at an aperture of f/22, to ensure sufficiently large depth of focus. 
Two of the cameras are mounted on each side of the test section, with optimal angles between adjacent cameras of $\sim 28-32^{\circ}$ for the particle triangulations, i.e. using a similar setup to our previous experiments in a 2-D contraction \citep{mugundhan2020alignment}.
A high-speed dual-cavity pulsed Nd-YLF green laser (527 nm) provides a beam which is expanded with two cylindrical lenses to illuminate a volume-slice about 23 mm wide.  The tracers are surfactant-treated fluorescent red polyethylene microspheres $\sim 50-70\; \mu$m (from Cospheric).  

The measurements were performed in three separate regions (P1, P2  and P3) along the length of the contraction, as shown in Fig. \ref{Water_Tunnel}(b). 
The corresponding volume slices were $106\times106\times23$ mm$^3$, $106\times56\times23$ mm$^3$ and $85\times38\times23$ mm$^3$.
To capture the particles in the accelerating flow, the frame-rate must be increased in the different regions within the contraction, increasing in the streamwise direction from 1000, 2000 to 4000 fps in volume P3 nearest to the outlet.

Following image pre-processing and spatial calibration with a dotted target,  
volume self-calibration is applied \citep{Wieneke2008}.
We track around 210,000, 80,000 and 50,000 particles in the differently sized regions. The Lagrangian particle-tracks are then mapped onto a regular Eulerian grid, of size $48\times48\times48$ voxels.  
With 75\% overlap we reconstruct 1 million, 600,000 and 300,000 velocity vectors within the P1, P2 \& P3 regions, respectively, with spacing of $\Delta x = 0.66$ mm.
We use time-filter length of $5$ for rms statistics or $11$ timesteps to improved \enquote{trackability} of the conherent vortical structures.

\section{Results} 

\begin{figure}
\begin{center}
\begin{tabular}{ c c c }
\includegraphics[width=0.33\textwidth]{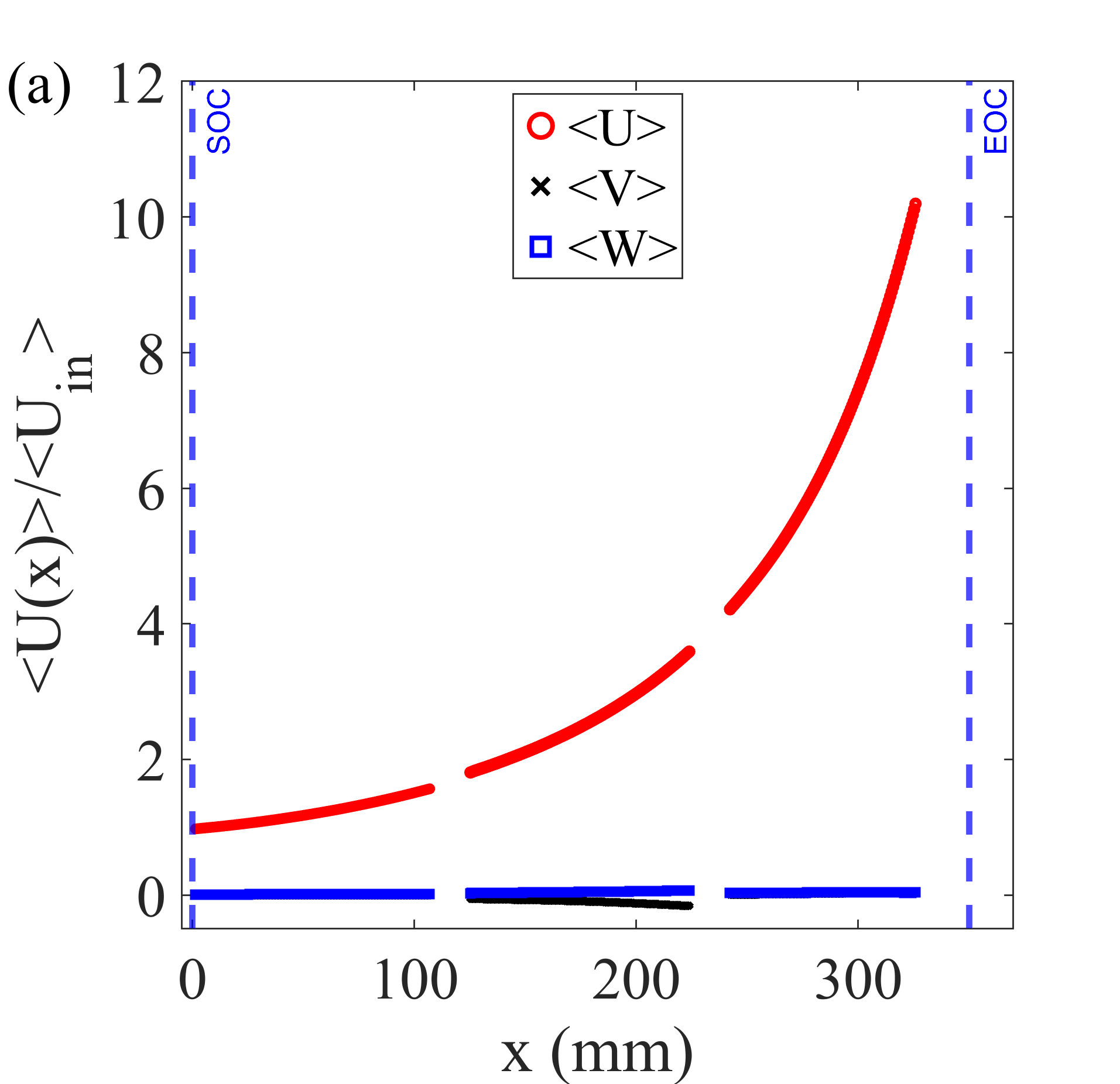} &
\includegraphics[width=0.33\textwidth]{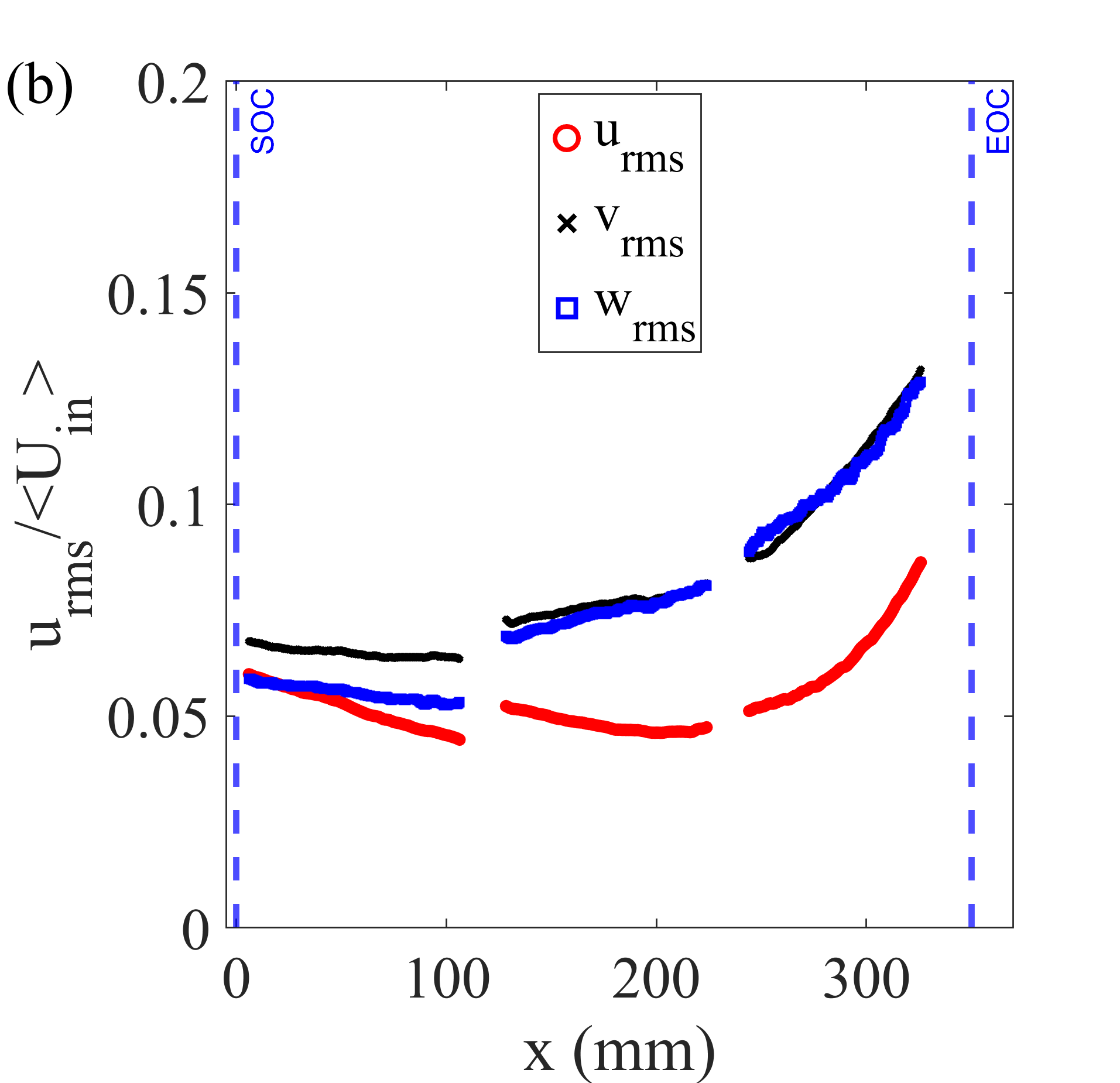}\vspace{-0.1in}&
\includegraphics[width=0.33\textwidth]{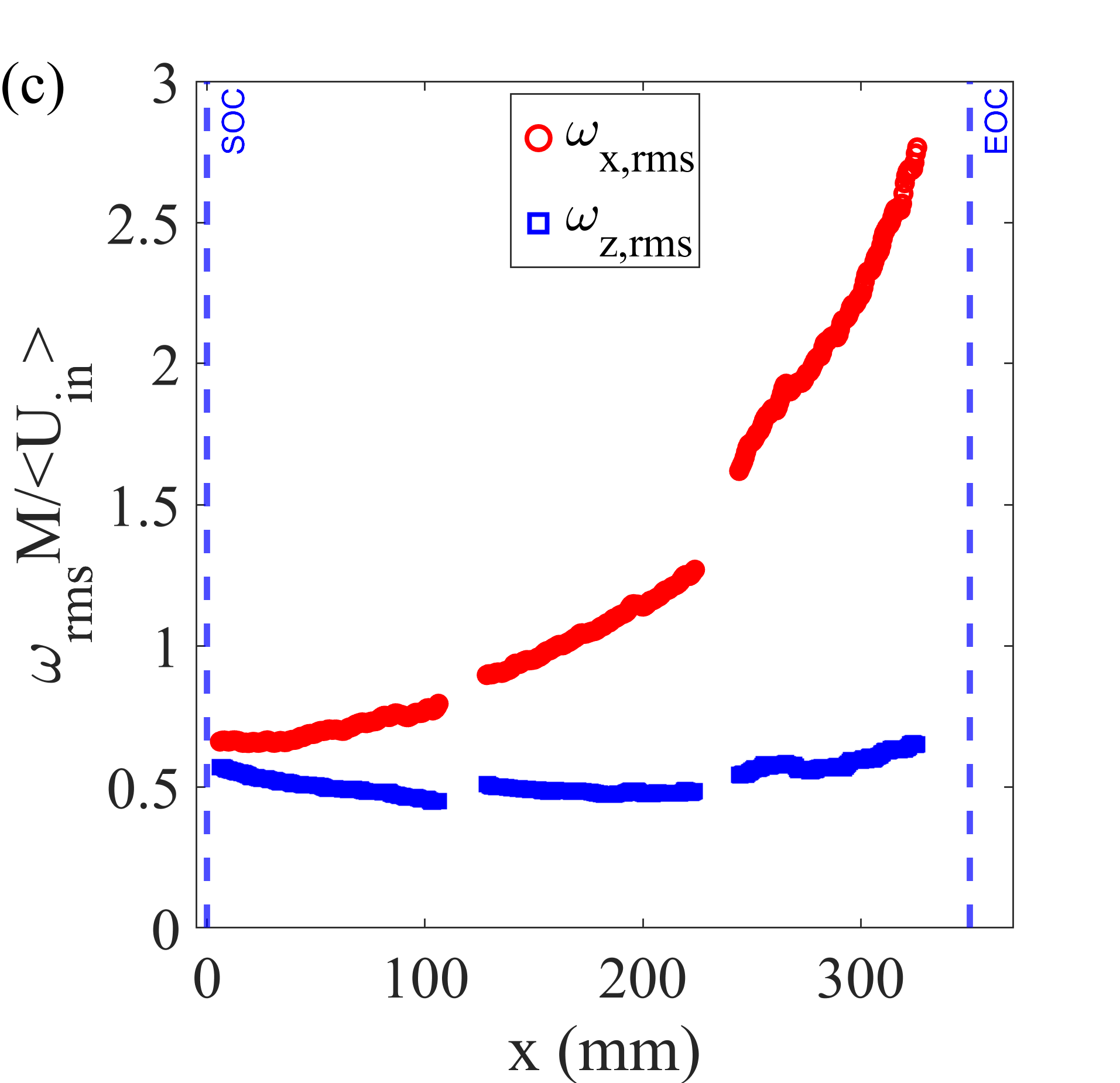}\vspace{-0.1in}\\
\end{tabular}
\end{center}
\caption{(a) Mean velocity components along the centerline of the contraction. 
(b) The r.m.s. velocity components normalized by the mean inlet velocity $\left\langle U_{\text{in}} \right\rangle$.
(c) Vorticty rms for the streamwise $\omega_{x,rms}$ and transverse component $\omega_{z,rms}$.} 
\label{MeanVelocity}
\end{figure}

\subsection{Mean velocity and fluctuation levels}

Figure \ref{MeanVelocity}(a) shows the mean streamwise velocity along the centerline of the contraction.  Near the exit, at the lowest part of the measurements volumes, the mean velocity has increased to $\sim 9$ times the inlet velocity, while the transverse mean velocities are near-zero as expected by symmetry.  The rms velocity components are shown in (b) and are consistent with earlier experiments, starting with suppression of the streamwise $u_{rms}$, while steady increase in the transverse fluctuations, owing to stretching of streamwise vortices.  In the lowest test volume $u_{rms}$ starts growing again, through non-linear interactions.  
Figure \ref{MeanVelocity}(c) shows the same scenario in terms of the vorticity components, where the streamwise $\omega_{x,rms}$ grows by a factor of 4, while the transverse component reduces slightly then stays about constant, with $\omega_{x,rms}/\omega_{z,rms} \simeq 4.5$, at the lowest measurement location in P3, 10 mm before the end of contraction, where the local $C\simeq 11$.  The vorticity fluctuations, which have not been measured previously, show results qualitatively consistent with the above basic inviscid theory, while the magnification is much below a factor of $C$.

\subsection{Coherent Vortical Structures}

\begin{figure}
\begin{center}
\begin{tabular}{ c c c c }
\includegraphics[width=0.24\textwidth]{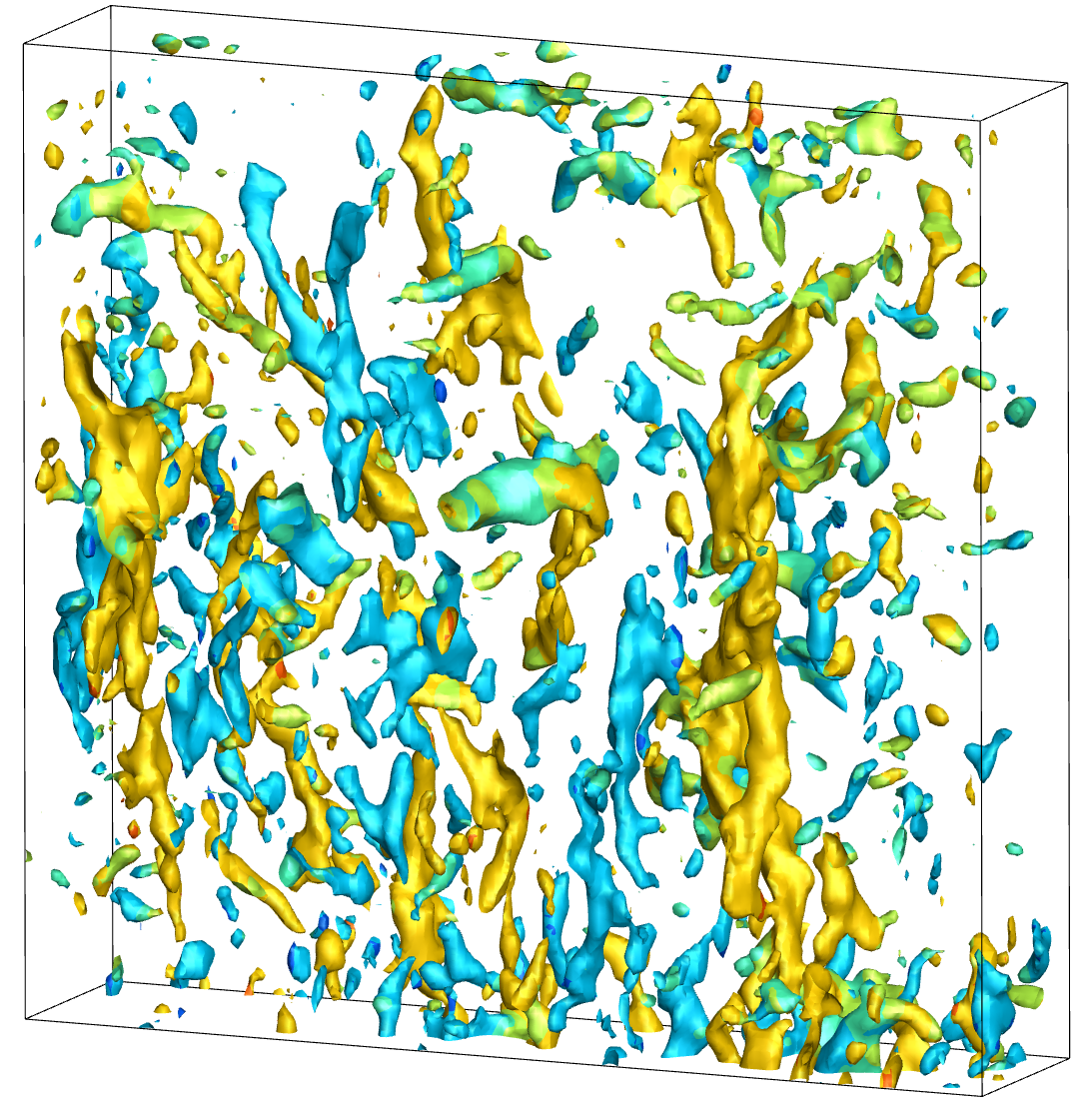} &
\includegraphics[width=0.24\textwidth]{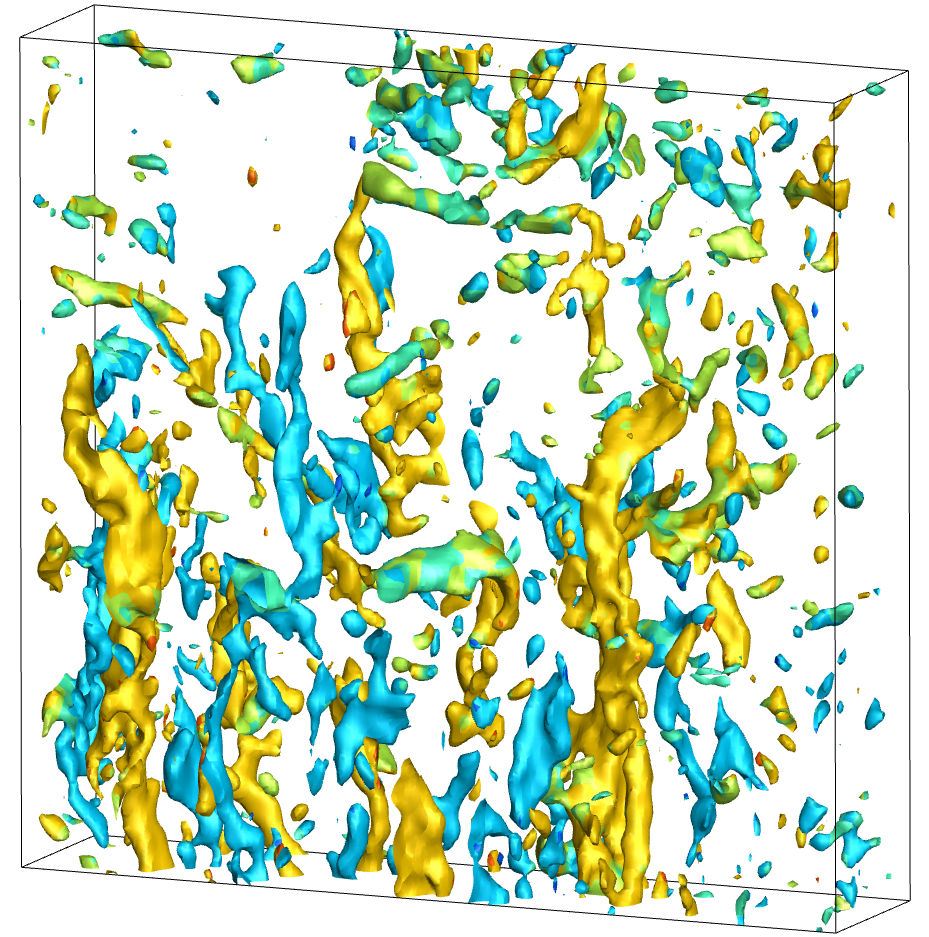} &
\includegraphics[width=0.24\textwidth]{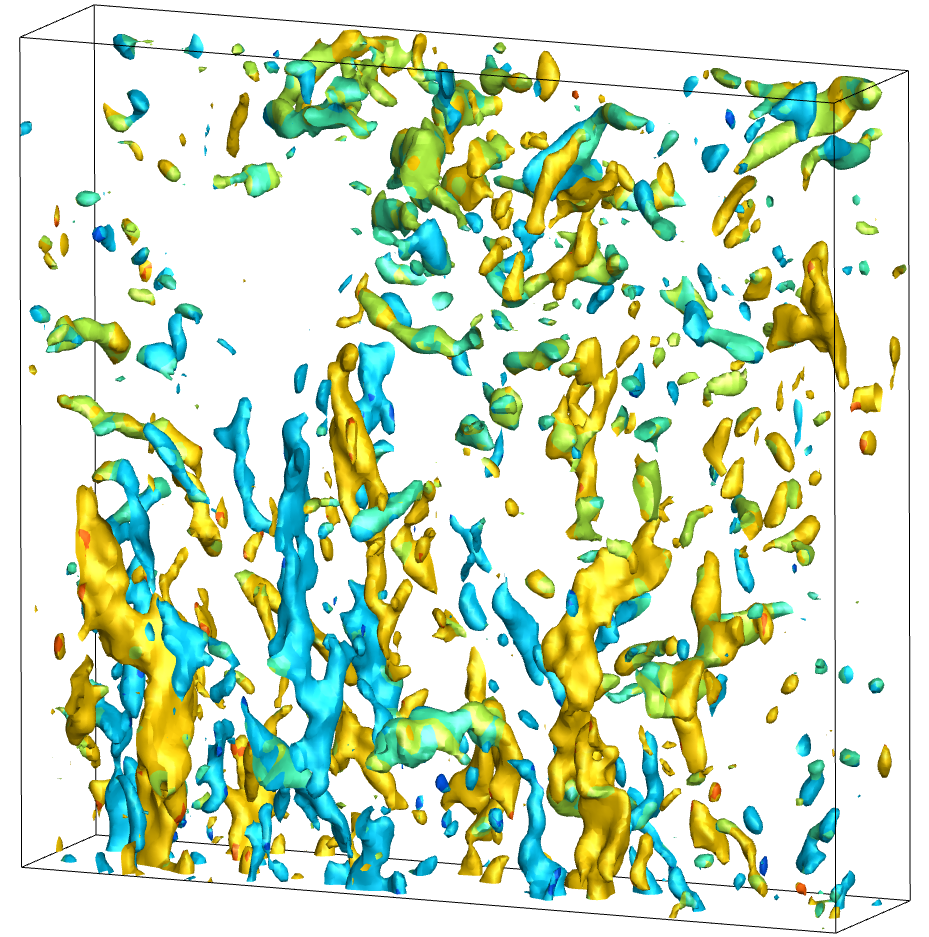} &
\includegraphics[width=0.24\textwidth]{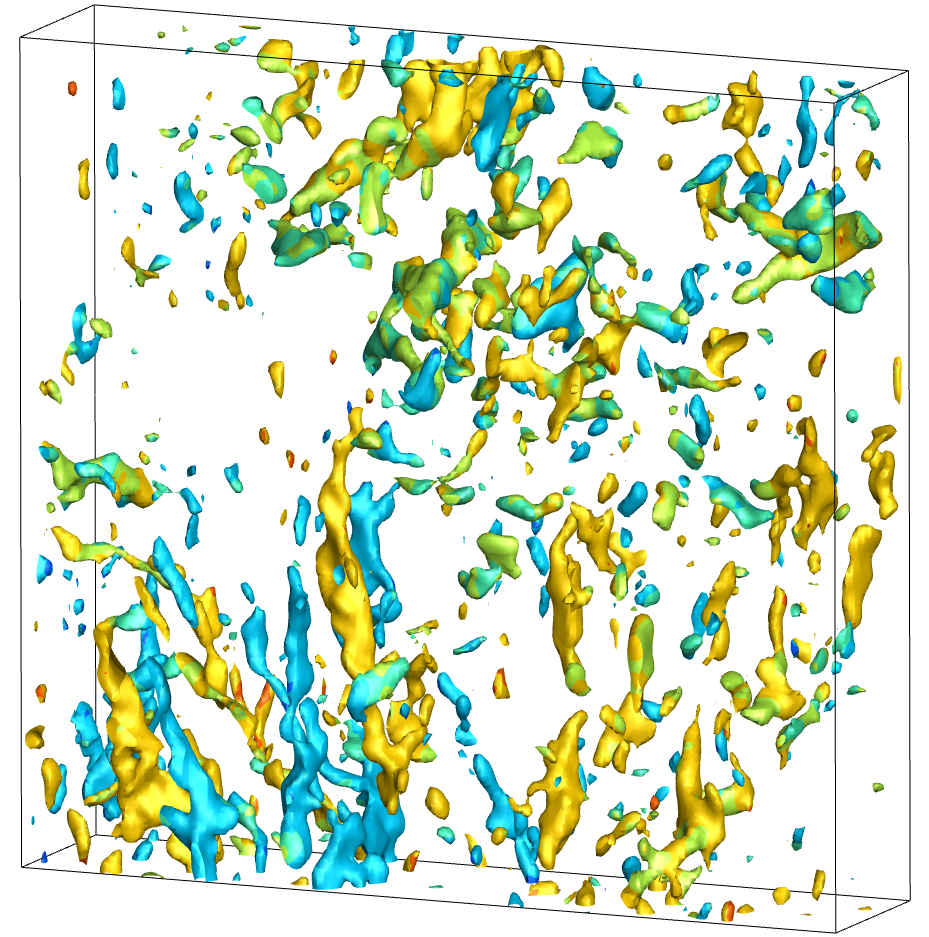} \\

\includegraphics[width=0.22\textwidth]{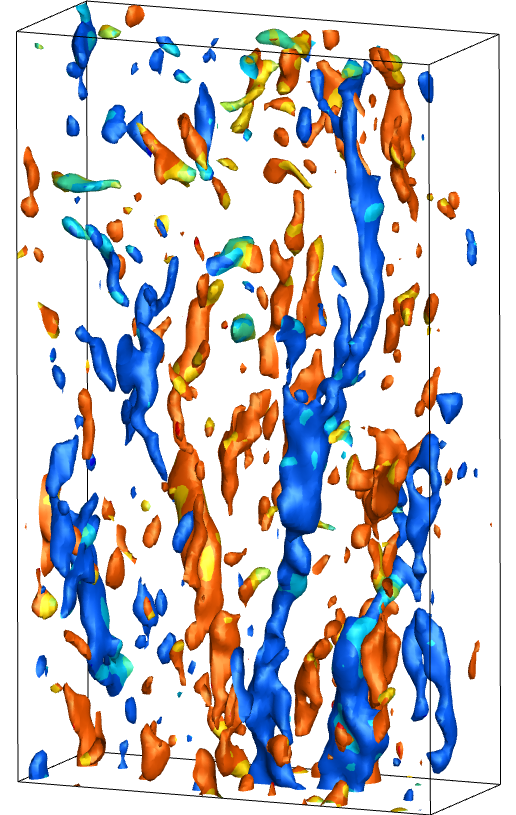} &
\includegraphics[width=0.22\textwidth]{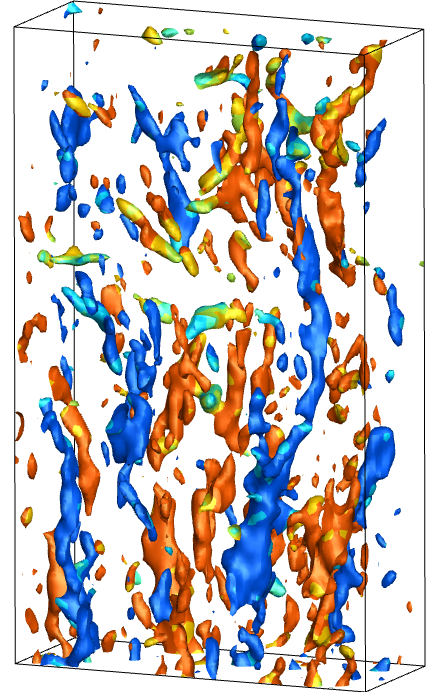} &
\includegraphics[width=0.22\textwidth]{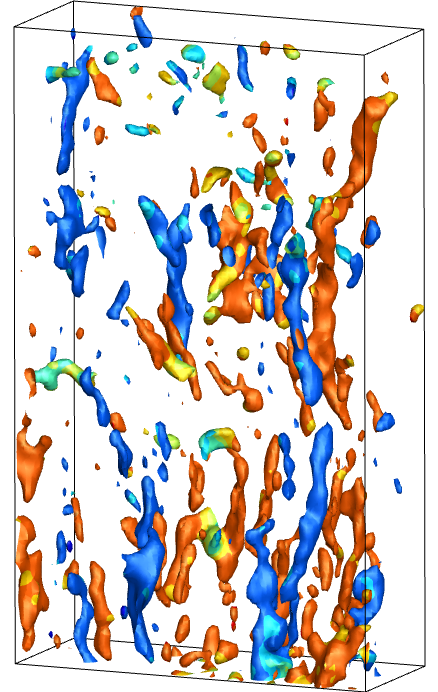} &
\includegraphics[width=0.22\textwidth]{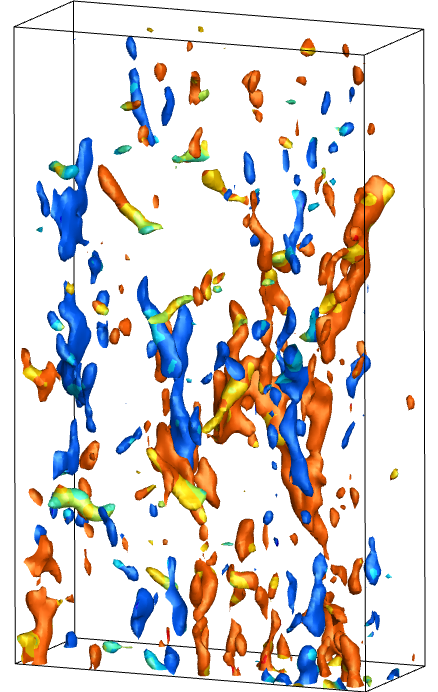} \\

\includegraphics[width=0.18\textwidth]{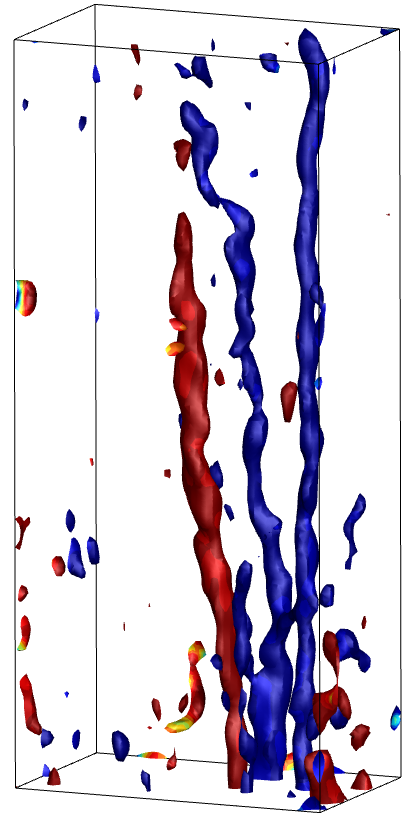} &
\includegraphics[width=0.18\textwidth]{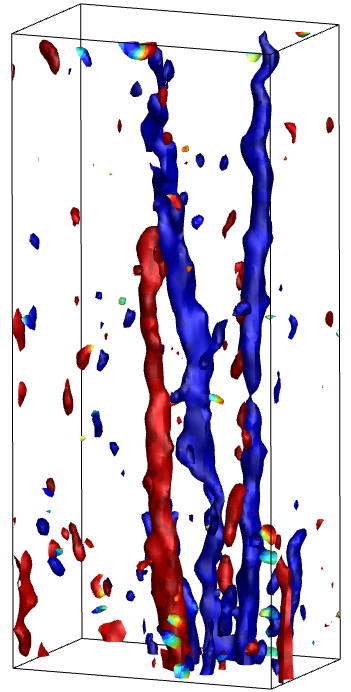} &
\includegraphics[width=0.18\textwidth]{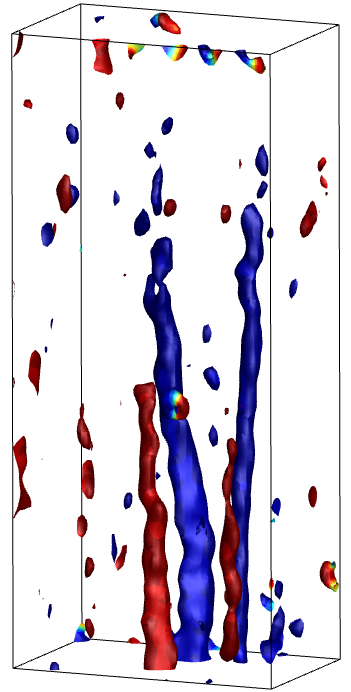} &
\includegraphics[width=0.18\textwidth]{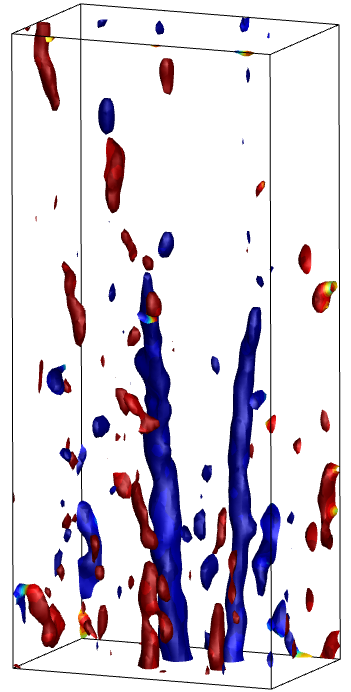} \\
\multicolumn{4}{c}{\includegraphics[width=0.9\textwidth]{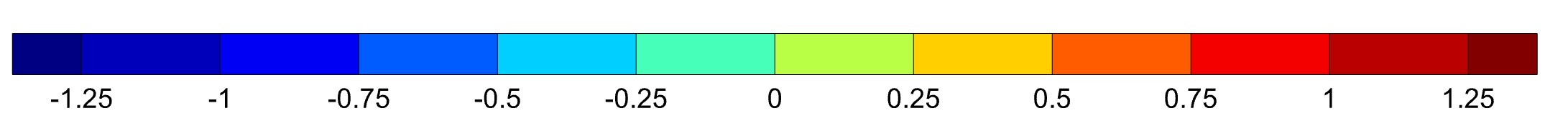}}\vspace{-0.1in}\\
\end{tabular}
\end{center}
\caption{The time evolution of the iso-surfaces of vorticity magnitude normalized by the maximum mean strain rate, ${\left | \omega \right |}/S^{}_{\text{max}}$, where the maximum measured value $S_{max} = |\partial \left \langle U \right \rangle/\partial x|_{max}$ = 32 s$^{-1}$. The iso-surfaces are colored based on the rotation direction of each structure with respect to the vertical, i.e. the sign of $\omega_{x}$. Each row shows a sequence corresponding to the different measurement regions, captured in different experimental runs. The volumes shown are separated by 60 video frames, captured at frame rates of 1000 fps for P1, 2000 fps for P2 and 4000 fps for P3. This corresponds to time intervals 60 ms, 30 ms and 15 ms, respectively. (a) Volume P1: $x = [0,106]$ mm, with threshold ${\left | \omega \right |}$ = 15 s$^{-1}$, or ${\left | \omega \right |}/S_{\text{max}} = 0.47$. (b) Volume P2: $x = [124-230]$ mm, with ${\left | \omega \right |}$ = 20 s$^{-1}$, or ${\left | \omega \right |}/S_{\text{max}} = 0.63$. (c) Volume P3: $x = [244-325]$ mm, with ${\left | \omega \right |}$ = 40 s$^{-1}$, or ${\left | \omega \right |}/S_{\text{max}} = 1.25$.  The depth of the volumes into the board are all 23 mm.  Each row is for a different experimental run.}
\label{Vortical_Structures}
\end{figure}

\begin{figure}
\begin{center}
\begin{tabular}{ c}
\includegraphics[width=0.8\textwidth]{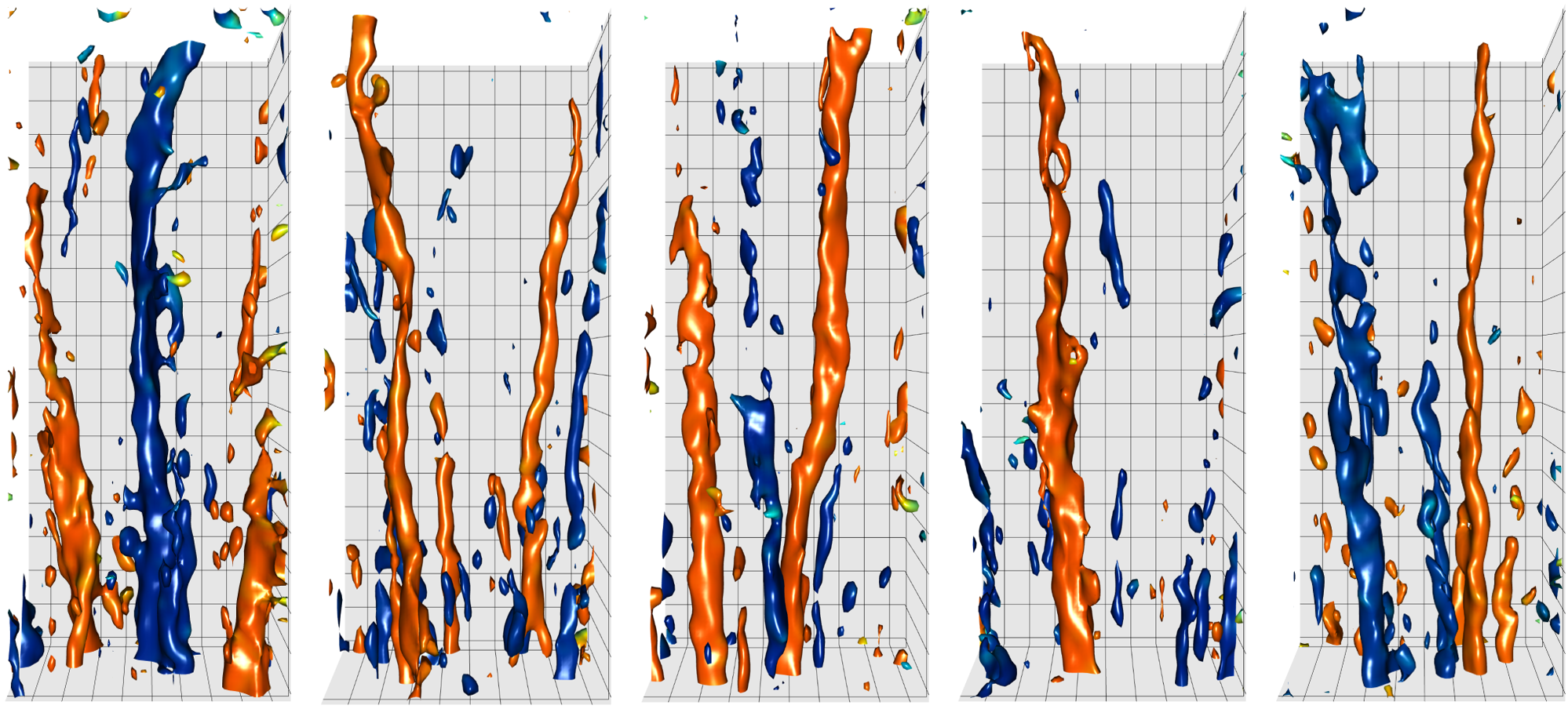}\vspace{-0.15in}\\
\end{tabular}
\end{center}
\caption{Five examples of the coherent vortical structures, each taken from the a different experimental run.  The isosurfaces correspond to $|\omega|=40$ s$^{-1}$ with the two colors indicating opposite sign rotation directions.
The supplemental material contains snapshots of all 795 observed structures.}
\label{Figure_4}
\end{figure}

Having acquired volumetric measurements we can, in addition to the rms fluctuations, investigate coherent vortical structures.  
Figure \ref{Vortical_Structures} shows examples of the vortical structures in the three measurement regions within the contraction, the locations of which were shown in Fig. \ref{Water_Tunnel}(b).  
The structures are identified by the isosurfaces of vorticity.
At the entrance to the contraction the vortices are small and fairly randomly oriented,
while in the center they become longer and a few of them are more prominent.
On the other hand, near the outlet, where the mean strain is largest, the flow is characterized by isolated long streamwise vortices.  
Furthermore, the strength of these vortices has been greatly enhanced towards the outlet.  
In these plots the ${\left | \omega \right |}-$threshold is increased in the streamwise direction from, 15, 20 to 40 s$^{-1}$. 
These coherent streamwise structures emerge repeatedly and dominate the observed turbulent structure.  
Figure \ref{Figure_4} shows five examples in the lowest volume P3, taken from five different experimental runs.

Using a ${\left | \omega \right |}$ threshold and a watershed algorithm we identify the coherent structures based on their volumes.
We also fit an ellipsoid around the largest vortices, which provides an estimate for their orientation and length. 
\textcolor{black}{The choice of ${\left | \omega \right |}$ threshold and the ellipsoid fitting is discussed in the Supplemental material.}   
Figure \ref{Fig_alignment}(b) shows the probability of vortex lengths $L$, which grows greatly as the flow is advected through the contraction.  Keep in mind that the threshold used in this calculation is increased as the vortices strengthen from regions P1, P2 to P3.  Furthermore, in volume P3 closest to the exit the vortices frequently span the entire length of the measurement volume (see Fig. \ref{Vortical_Structures}c) and $L$ is therefore a lower bound for the longest vortices.

The extended time-series of volumes allows us to quantify the prevalence of these coherent vortical structures.  We identify them by looking at the time-series of the longest structures within each volume in the bottom section $P3$.
Section of a typical time-series of vortex-structure lengths $L$ is shown in Fig. \ref{Fig_alignment}(a).  
Over 5 separate experimental runs, we identified 795 structures, or on average $159 \pm 20$ separate structures per each continuous experimental run of 136,502 volumes. 
\textcolor{black}{One characteristic structure from each run is shown in Figure \ref{Figure_4}.} 
The {\it Supplemental Material} includes a snapshot of one volume from each of the 795 long vortices, to give an indication of the variability in their morphology.  Each structure is advected through the volume and is visible in $\sim 600 \pm 75$  volumes, corresponding to 150 ms.  
For comparison this time is 2.5 times the advection time-scale computed by $T = \int_{t_{1}}^{t_{2}} dt = \int_{x_{1}}^{x_{2}} \left\langle U(x) \right\rangle^{-1} dx = 60.5$ ms, which marks how long it takes a fluid element to be advected through the P3 volume at the mean local velocity. This hints at the vortices being longer than the height of P3, as one should arrive at a time of $2T$ for a vortex of the same length as the volume, counting from the bottom tip of the vortex entering the volume until the top of the vortex leaves the bottom.

The large number of structures investigated also allows us to ascertain the symmetry of the turbulence in our experimental tunnel.  The number of clockwise and counter-clockwise coherent vortices should be identical for a perfectly symmetric inlet conditions.  This is indeed observed with 49/51 \% of the structures with each sign of the rotation direction.  This suggests that the grid rotation protocol and the horizontal location of the grid rods do not bias the formation of the structures towards one sign and rules out large-scale rotations in the incoming flow.

\begin{figure}
\centering
\begin{tabular}{ c c}
\includegraphics[width=0.385\textwidth]{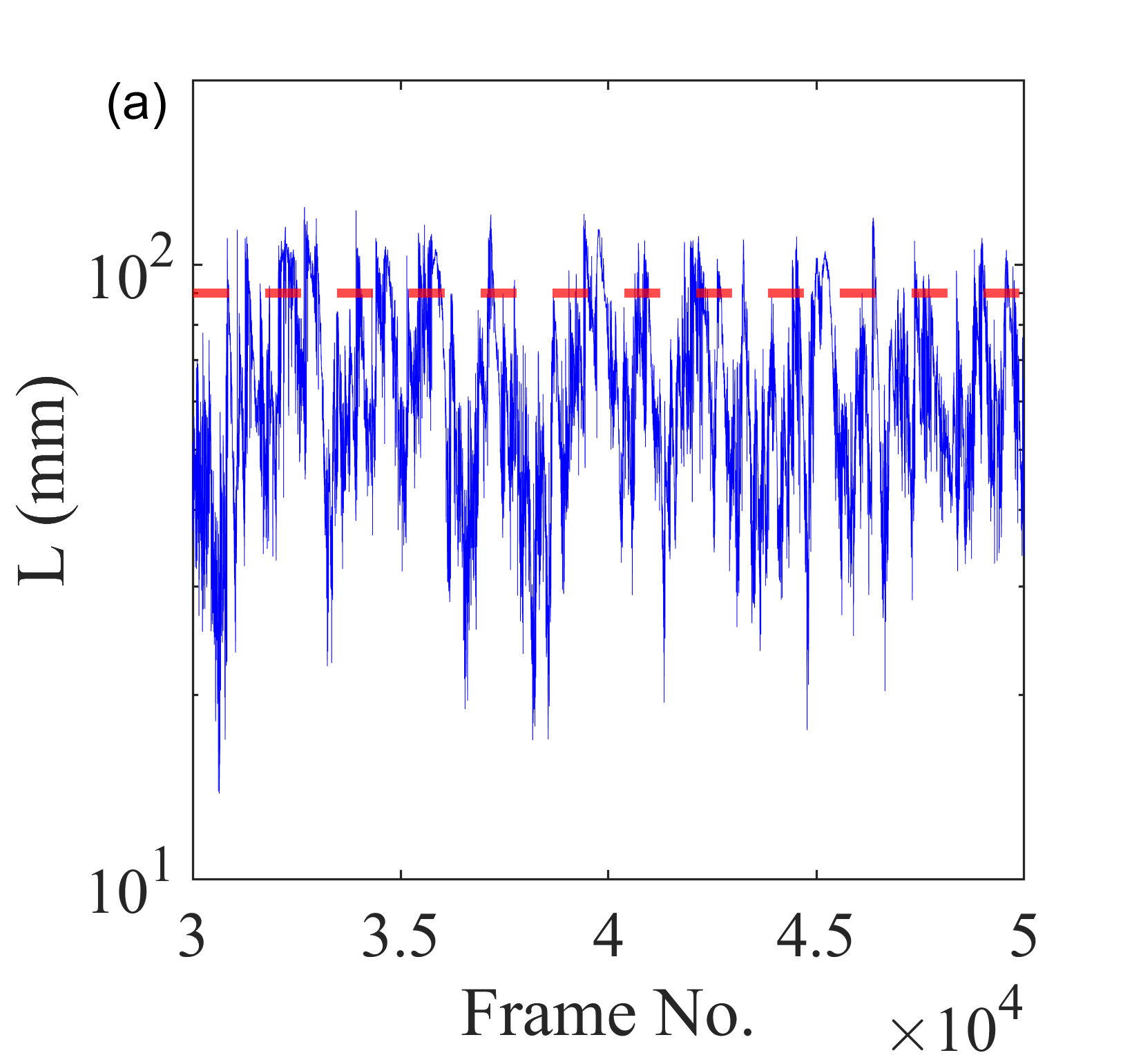} &
\includegraphics[width=0.38\textwidth]{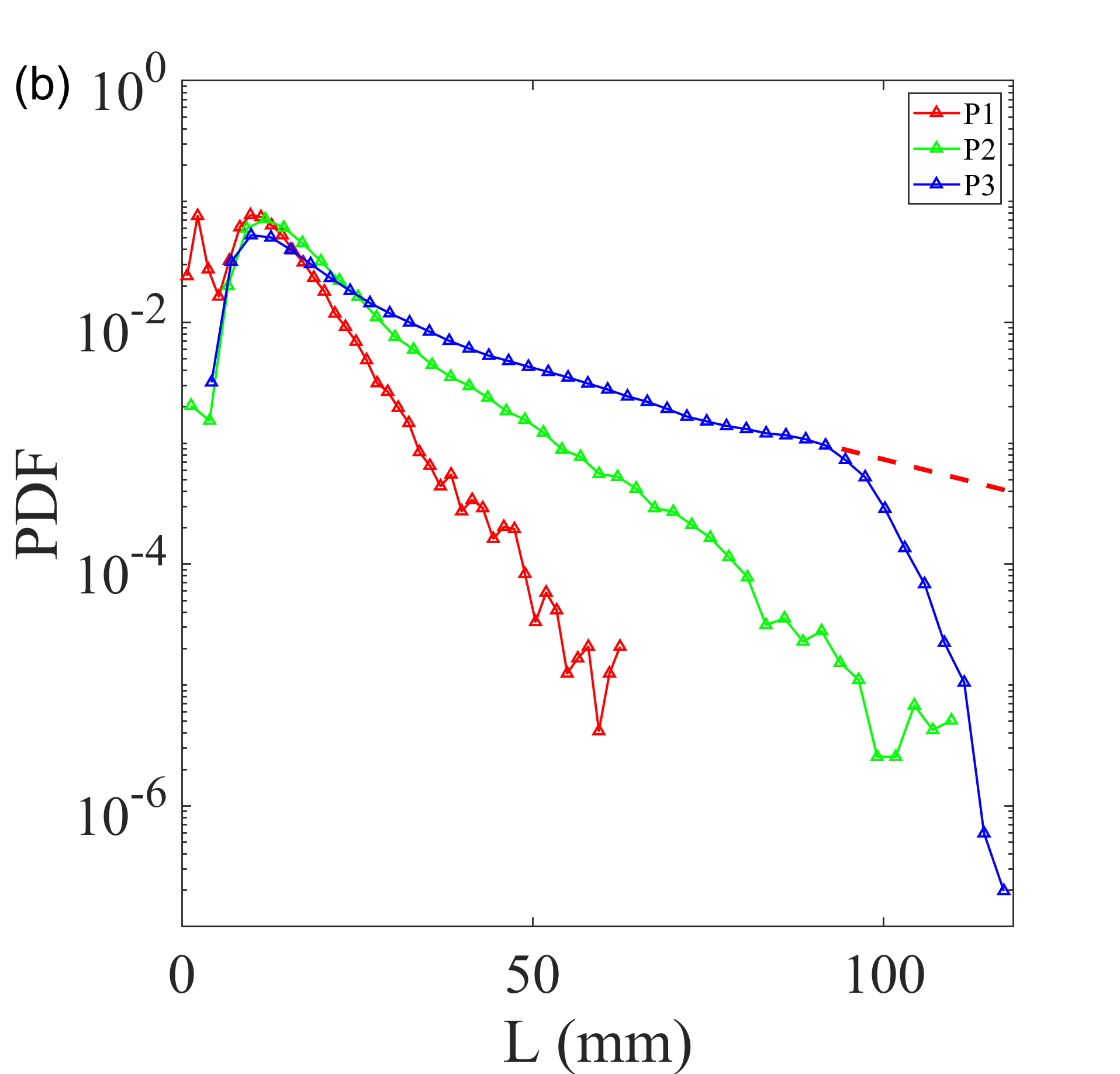}\\
\includegraphics[width=0.36\textwidth]{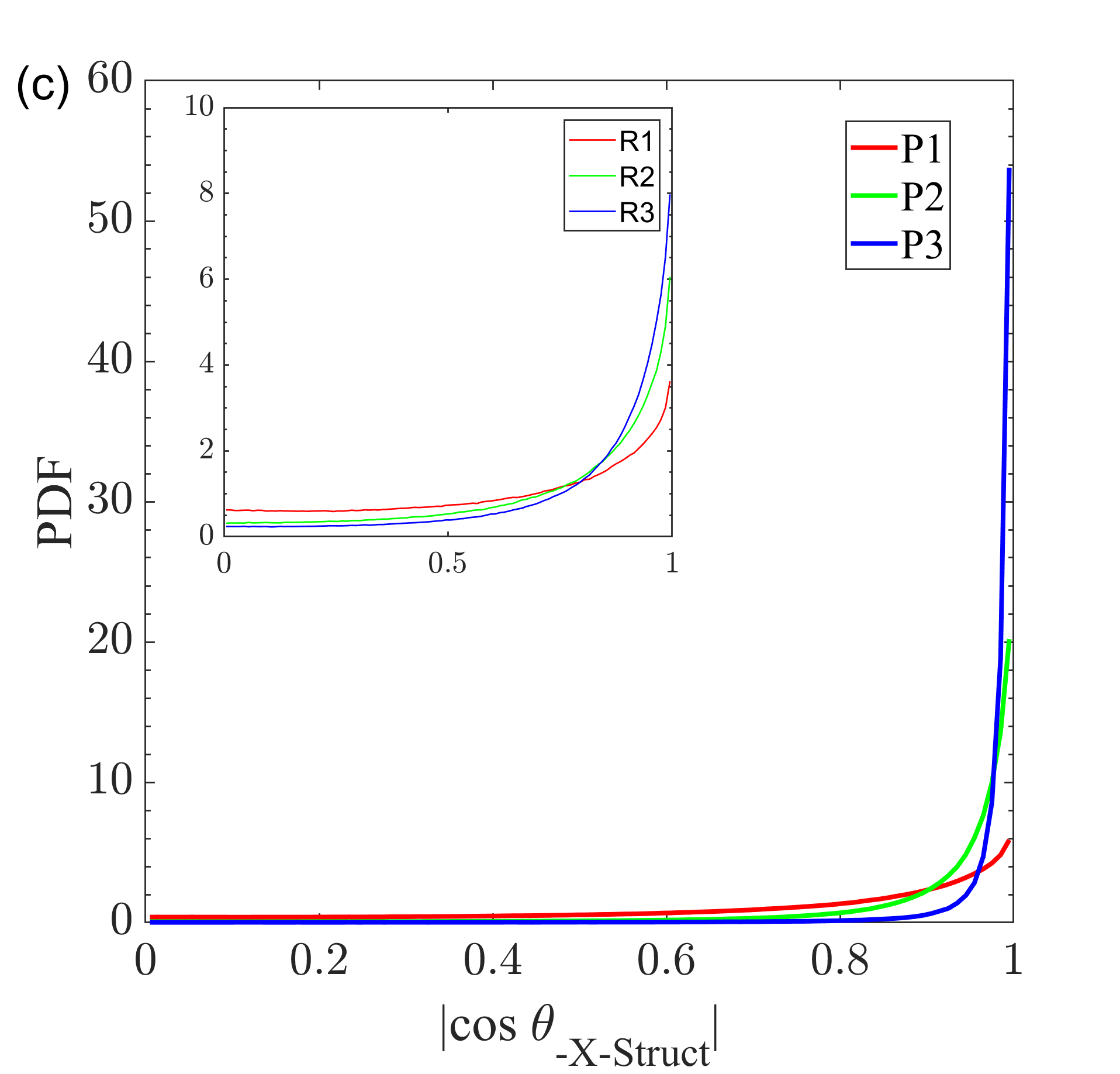}\vspace{-0.25in}&
\includegraphics[width=0.36\textwidth]{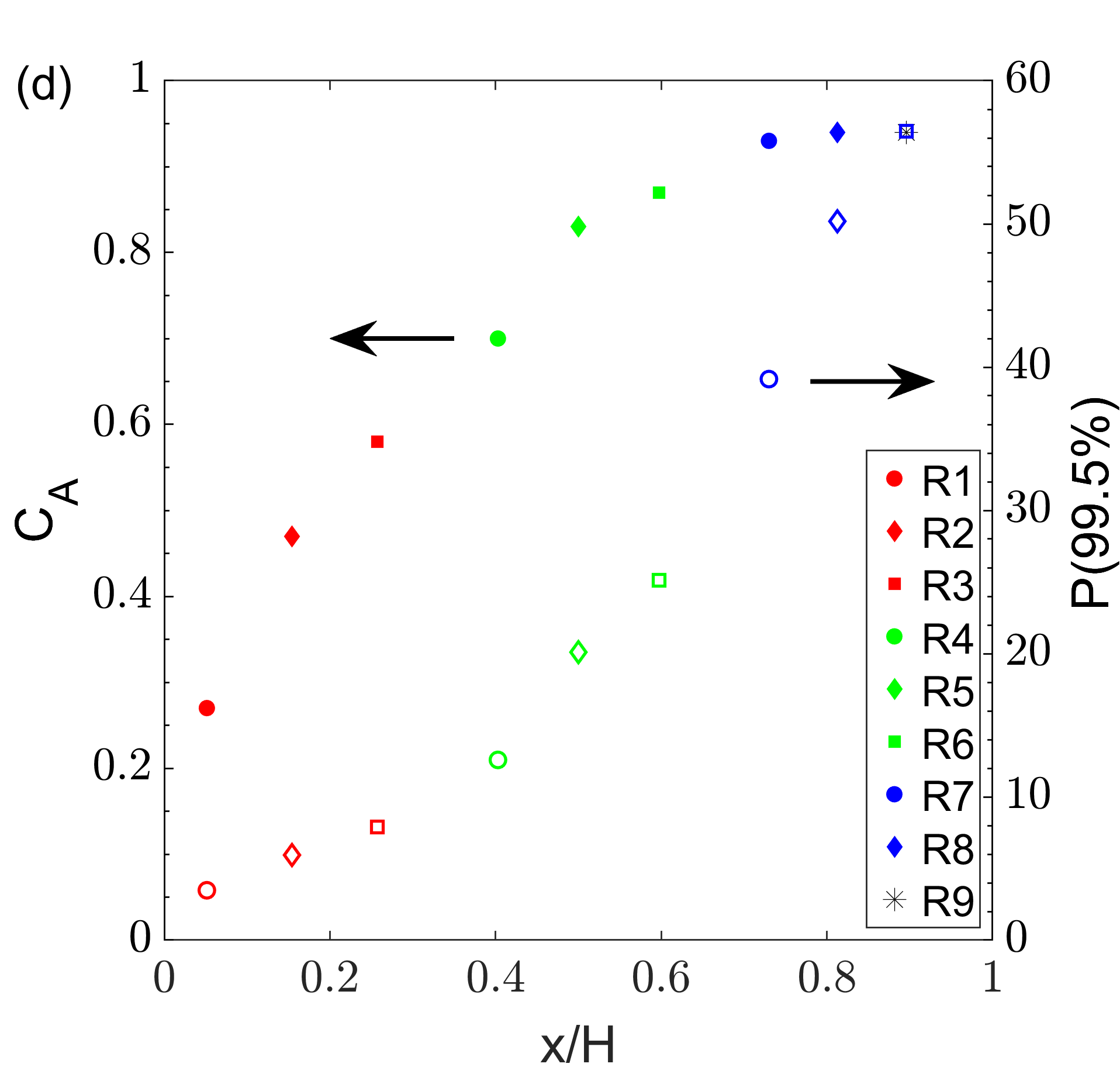}\vspace{-0.1in}\\
\caption{Length and alignment of the coherent vortical structures.
(a) Time-series of the length of the longest vortical structure within the volume P3 near the exit of the contraction.  When the curve exceeds the horizontal line of \textcolor{black}{$L=90$} mm, the vortex spans the entire height of the volume. 
(b) Probability of the length of the coherent structures, for the three measurement regions.  
\textcolor{black}{The dashed red curve is an approximate extrapolation to highlight that some of the vortices span the entire length of section P3 and must thereby be longer than the volume, due to clipping.}
(c) Streamwise evolution of the PDF of the orientation of coherent vortical structures.  Only structures larger than 100 voxels are included, while the $\boldsymbol {\left | \omega  \right |}-$threshold in the three locations are varied: in region P1 $\boldsymbol {\left | \omega  \right |}$= 15 s$^{-1}$; in P2: $\boldsymbol {\left | \omega  \right |}$= 20 s$^{-1}$ and in P3: $\boldsymbol {\left | \omega  \right |}$= 40 s$^{-1}$. The inset plot highlights the orientation a short distance inside the contraction, where the orientation is closest to uniform, in subregions of P1, with streamwise ranges of: R1, $x=0-35.3$ mm; R2, $x=35.5-70.7$ mm and R3, $x=70.7-106$ mm.
(d) The prevalence of streamwise orientation of the coherent structures along the contraction, quantified by the orientation coefficient $C_A$ (solid symbols) and peak value at 99.5\% probability (open symbols). The red, green and blue symbols correspond to measurements in volumes P1, P2 and P3, respectively.  The streamwise coordinates from the entrance to the contraction are 18, 54, 90, 141, 175, 209, 256, 284 and 314 mm, which are normalized by the total length of the contraction  $H=350$ mm.
\textcolor{black}{The arrows point to which axes correspond to the different data sets.}}
\label{Fig_alignment}
\end{tabular}
\end{figure}

\subsection{Vortex alignment}

The coherent structures are best described by their extensive length and alignment with the streamwise mean strain.
Figure \ref{Fig_alignment}(c) shows how strongly the coherent vortices become aligned with the mean strain, with the peak in the probability concentrating at the angle $\theta = 0$, i.e. $cos\theta = 1$.  However, the height of the peak of the pdf depends on the number of bins used, i.e. how well it is resolved near $cos\theta = 1$.  
We calculate the PDf using $cos\theta$ to account for the range of allowed azimuthal space for each $\theta$.
We propose here two quantitative measures independent of the bin-size:  First, what is the value of the probability when 99.5\% of cumulative distribution has been reached, starting at $cos\theta = 0$.  This is found by fitting the pdfs with an inverted fifth-order polynomial (see Supplemental Material for details).  This value is plotted in Fig. \ref{Fig_alignment}(d).  Secondly, we form an orientation coefficient $C_A$,
which takes a value of zero for uniform orientation and unity at fully aligned orientation, which is defined as twice the moment of the probability around $cos\theta = 0.5$:
\[
C_A = \int_0^{\pi/2} 2\, [ cos\,\theta - 0.5 ] \; P( cos\,\theta ) \; d\theta.
\]
Keep in mind that we have multiplied by a factor of 2, to confine possible values of the orientation coefficient in the range $C_A \rightarrow [-1,1]$.
Figure \ref{Fig_alignment}(d) shows the value of this coefficient and how it approaches unity traveling through the contraction.

\subsection{Vortex stretching}

\begin{figure} 
\centering
\begin{tabular}{ c c}
\includegraphics[width=0.38\textwidth]{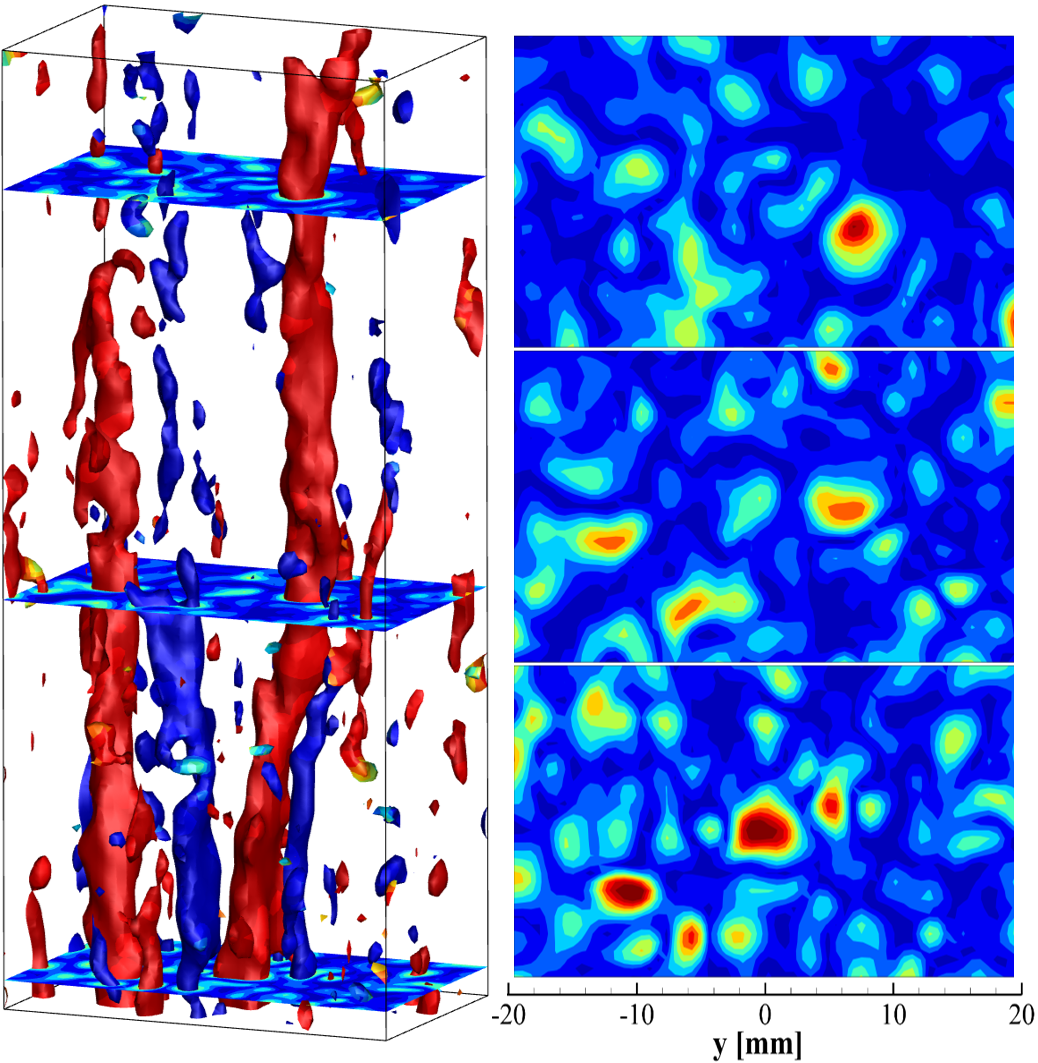} &
\includegraphics[width=0.42\textwidth]{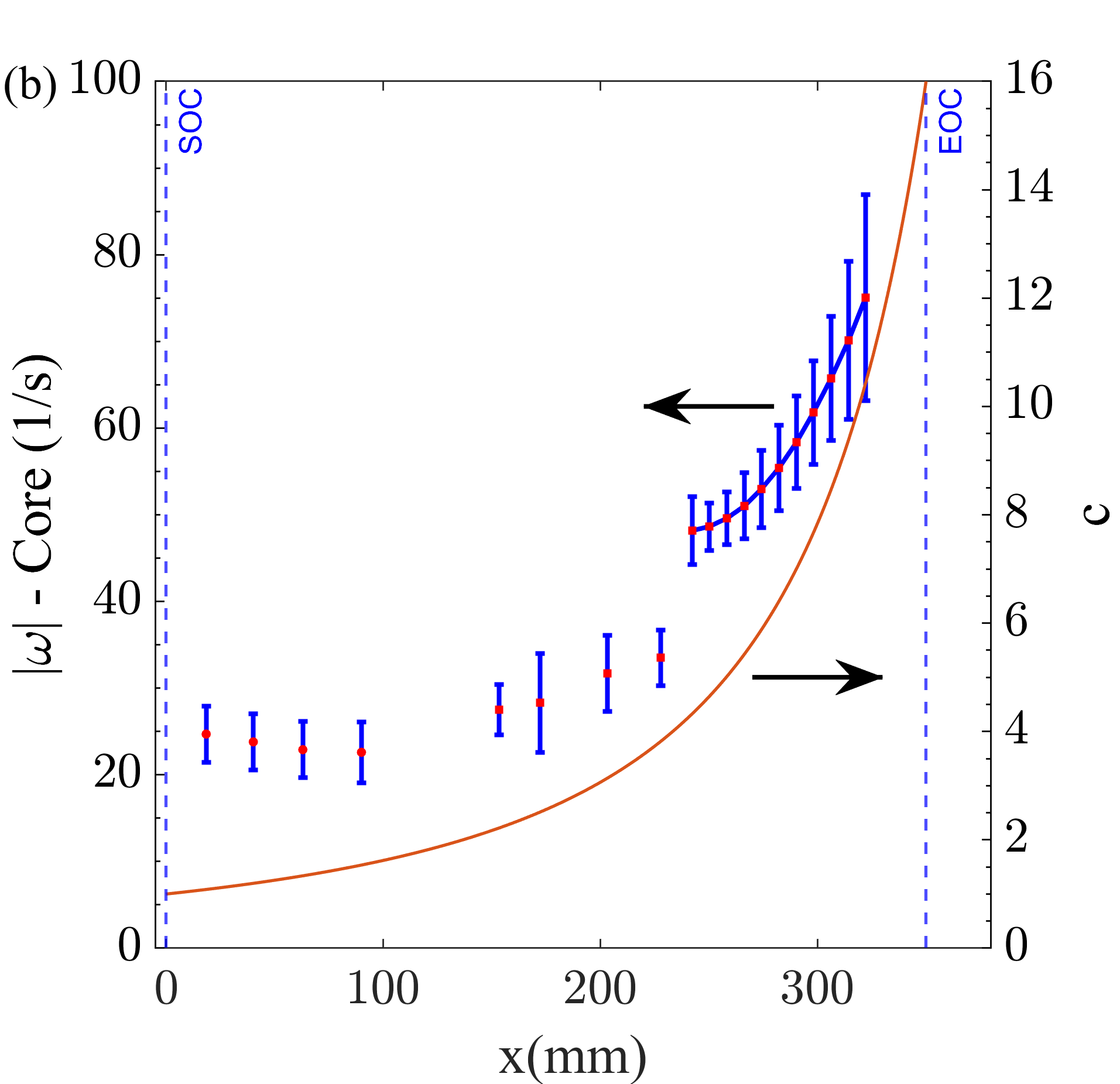}\vspace{-0.25in}\\
\caption{(a) Instantaneous vortical structure visualized in region P3 with threshold $\boldsymbol {\left | \omega  \right |}$= 40 s$^{-1}$, with three corresponding horizontal cuts through the vortex cores, showing their vorticity. 
(b) The streamwise variation of $\boldsymbol {\left | \omega  \right |}$ in the cores of typical coherent vortices. In regions P1 and P2, the core strength is extracted from about 24 structures in each region.
In P3 the vorticity is tracked along the longest coherent structures extending the entire region, with the mean of a quadratic fit obtained from 48 strong structures.
The red line indicates the local contraction ratio, $c$, shown on the right ordinate.
\textcolor{black}{The arrows point to which axes correspond to the curves.}}
\label{Stretching}
\end{tabular}
\end{figure}

Vortices aligned with the mean strain should experience an increase in the local vorticity strength along their length, due to the stretching and near-constancy of circulation.  This is verified in Fig. \ref{Stretching}(b), where the peak vorticity along the coherent vortices in P3, increases on average by about 65\%, but much smaller than the increase in $C$.
The vortex strengths are also significantly larger than near the inlet, where the vortices are more randomly oriented.  Figure \ref{Stretching}(a) shows that the diameters of the typical coherent vortices are $\simeq 3-6$ mm. The above core sizes are determined by the equivalent cross-section of the structures above the cut-off $\boldsymbol {\left | \omega  \right |} \geq 40$ s$^{-1}$ and their full size is $\sim 50$\% larger.

\begin{figure} 
\centering
\begin{tabular}{ c c}
\includegraphics[width=0.52\textwidth]{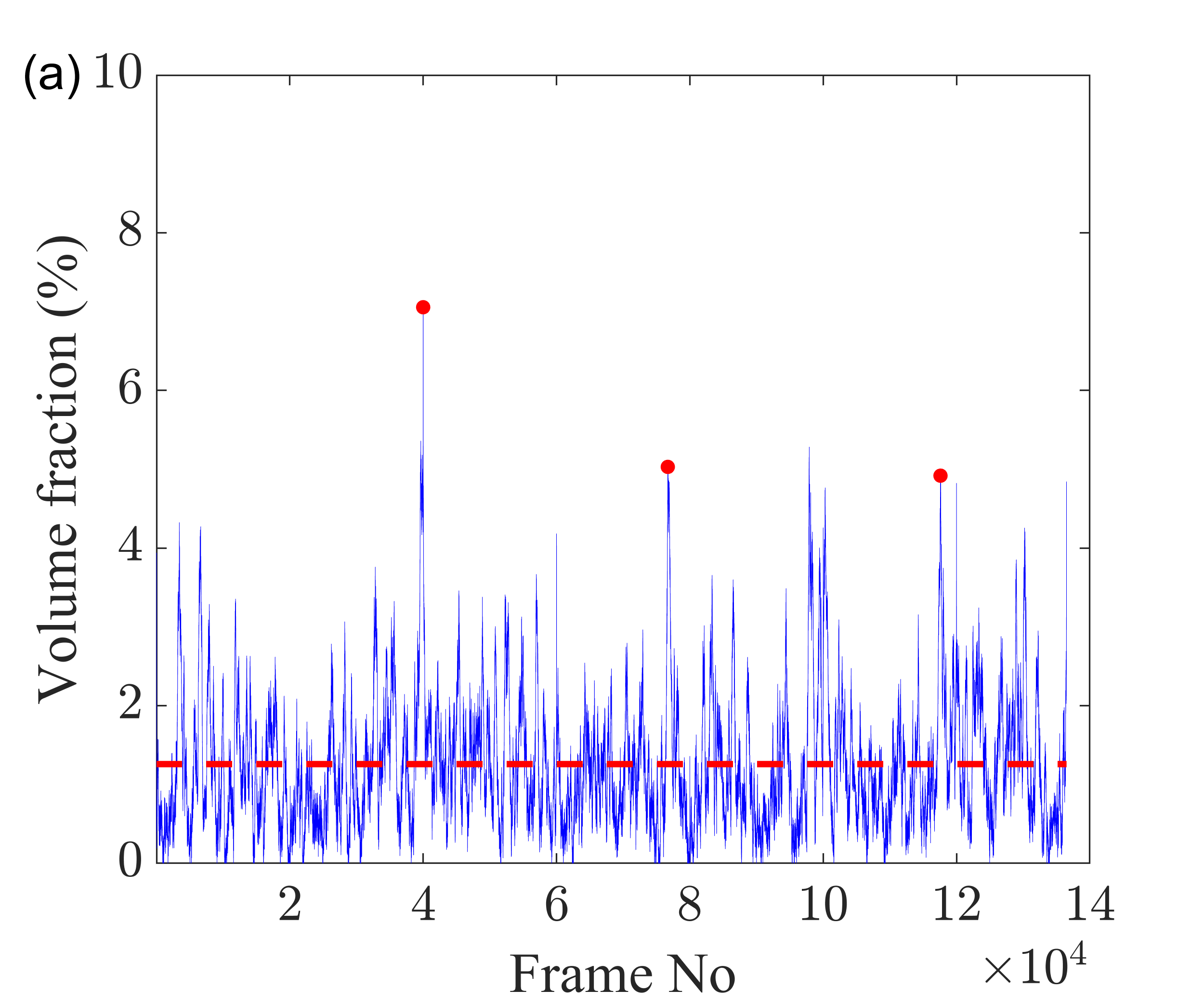} &
\includegraphics[width=0.45\textwidth]{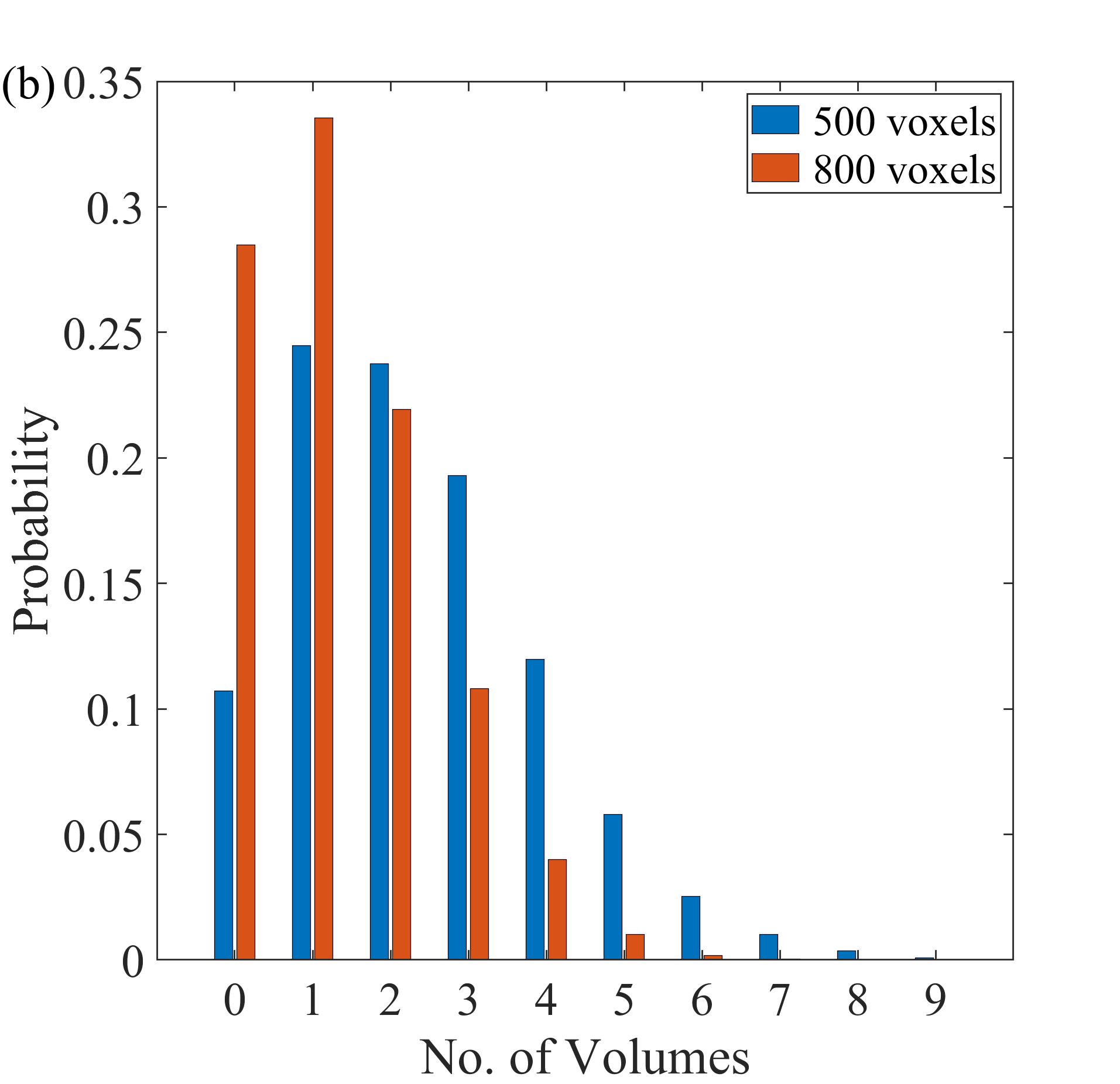}\vspace{-0.25in}\\
\caption{\textcolor{black}{(a) The time-series of the volume fraction, $V_{struct}/V_{total}$, of the coherent vortical structures in region P3 for one complete experimental realization - time between frames $dt=0.25$ ms. 
Here, $V_{struct}$ is the total volume of all structures identified by $\boldsymbol {\left | \omega  \right |} \geq 40$ s$^{-1}$ and having a volume greater than 300 voxels, while $V_{total}$ is the total volume of the measurement region. 
The instantaneous coherent structures corresponding to the maximum volume fraction values marked by red dots, are shown in Supplemental Figure S2. (b) The histogram of the number of coherent structures observed in each volume, for two different thresholds of structure volumes, of 500 \& 800 voxels (144 \& 230 mm$^3$), corresponding to 0.17\% and 0.27\% of $V_{total}$.}}
\label{Volume}
\end{tabular}
\end{figure}

\section{Discussion and Conclusions}

Our volumetric time-resolved vorticity measurements have identified prominent coherent streamwise vortical structures dominating the turbulence in a rapid contraction.
Even though these coherent structures can be anticipated from the basic conceptual picture of vortex stretching by the streamwise mean strain,
their strength, extensive length and prevalence has not been characterized before.  The coherent vortices frequently span the entire length of the imaged volume P3, which is approximately four times the integral scale of the incoming flow and about 14 times the Taylor microscale.

We propose two ways of characterizing the alignment of the structures in the streamwise, i.e. P99.5\% and $C_A$.  Both measures verify this strong alignment.

The typical diameter of the structures $\sim 7$ mm, is about an order of magnitude larger than the characteristic size of the Burgers vortex $D=\sqrt{4\nu/S} = 0.8$ mm, where viscous diffusion of vorticity is balanced by the enhancement by axial stretching, showing they are not viscous controlled.   The streamwise velocity inside the vortex cores is only $\sim 1$\% slower than the mean, thereby not contributing much to the $u_{rms}$ fluctuations.
The equal number of clockwise and counter-clockwise coherent vortices suggests they are not driven by large-scale rotation in the inlet flow into the tunnel.

Even though the coherent structure are the prominent vortical features in the flow, their volume is minimal.  
Using watershed algorithm we measure the volume of coherent structures with $\boldsymbol {\left | \omega  \right |} \geq 40$ s$^{-1}$ in each volume.
Figure \ref{Volume}(a) shows how this volume fraction changes with time, but looking at only the longest structures they instantaneously occupy $\simeq 0.5\%$ of the total measurement volume in P3.  Figure \ref{Volume}(b) shows the number of these structures likely to occupy a random volume.  It is more likely to have 1 or 2 of these structures than none, destroying any notion of small-scale homogeneity.

Future simultaneous measurements with multiple Tomographic systems could identify the progenitors of these coherent structures and how the multiple of vortices are winnowed down to one or two near the outlet.  
It also remains an open question whether these structures will persist at higher $Re_{\lambda}$.

\begin{acknowledgements}
{This study was supported by King Abdullah University of Science and Technology
(KAUST) under BAS/1/1352-01-01.}\\  
{\bf Declaration of Interests. The authors report no conflict of interest.}
\end{acknowledgements}


\bibliography{Short}

\begin{thebibliography}{31}
\expandafter\ifx\csname natexlab\endcsname\relax\def\natexlab#1{#1}\fi
\def\au#1{#1} \def\ed#1{#1} \def\yr#1{#1}\def\at#1{#1}\def\jt#1{\textit{#1}} \def\bt#1{#1}\def\bvol#1{\textbf{#1}} \def\vol#1{#1} \def\pg#1{#1} \def\publ#1{#1}\def\arxiv#1{#1}\def\org#1{#1}\def\st#1{\textit{#1}}

\bibitem[Ayyalasomayajula \& Warhaft(2006)]{Ayyalasomayajula2006}
{\sc \au{Ayyalasomayajula, S.} \& \au{Warhaft, Z.}} \yr{2006}  \at{Nonlinear interactions in strained axisymmetric high-reynolds-number turbulence}.  \jt{J. Fluid Mech.}  \bvol{566},  \pg{273--307}.

\bibitem[Batchelor \& Proudman(1954)]{Batchelor1954}
{\sc \au{Batchelor, G.~K.} \& \au{Proudman, I.}} \yr{1954}  \at{The effect of rapid distortion of a fluid in turbulent motion.}  \jt{Quart. J. Mech. Appl. Math.}  \bvol{7},  \pg{83--103}.

\bibitem[Brown \& Roshko(1974)]{BrownRoshko1974}
{\sc \au{Brown, G.~L.} \& \au{Roshko, A.}} \yr{1974}  \at{On density effects and large structure in turbulent mixing layers}.  \jt{J. Fluid Mech.}  \bvol{64},  \pg{775--816}.

\bibitem[Elsinga {\em et~al.\/}(2006)Elsinga, Scarano, Wieneke \& van Oudheusden]{Elsinga2006}
{\sc \au{Elsinga, G.~E.}, \au{Scarano, F.}, \au{Wieneke, B.} \& \au{van Oudheusden, B.~W.}} \yr{2006}  \at{Tomographic particle image velocimetry}.  \jt{Exp. Fluids}  \bvol{41}~(6),  \pg{933--947}.

\bibitem[Ertun\c{c} \& Durst(2008)]{Ertunc2008}
{\sc \au{Ertun\c{c}, "{O}.} \& \au{Durst, F.}} \yr{2008}  \at{Evidence of very long meandering features in the logarithmic region of turbulent boundary layers}.  \jt{Phys. Fluids}  \bvol{20},  \pg{025103}.

\bibitem[Ganapathisubramani {\em et~al.\/}(2003)Ganapathisubramani, Longmire \& Marusic]{Ganapathisubramani2003}
{\sc \au{Ganapathisubramani, B.}, \au{Longmire, E.~K.} \& \au{Marusic, I.}} \yr{2003}  \at{Characteristics of vortex packets in turbulent boundary layers}.  \jt{J. Fluid Mech.}  \bvol{478},  \pg{35--46}.

\bibitem[Grossmann {\em et~al.\/}(2016)Grossmann, Lohse \& Sun]{Grossmann2016}
{\sc \au{Grossmann, S.}, \au{Lohse, D.} \& \au{Sun, C.}} \yr{2016}  \at{High–{R}eynolds number {T}aylor-{C}ouette turbulence}.  \jt{Annu. Rev. Fluid Mech.}  \bvol{48},  \pg{53--80}.

\bibitem[Hamilton {\em et~al.\/}(1995)Hamilton, Kim \& Waleffe]{Hamilton_Kim_Waleffe_1995}
{\sc \au{Hamilton, J.~M.}, \au{Kim, J.} \& \au{Waleffe, F.}} \yr{1995}  \at{Regeneration mechanisms of near-wall turbulence structures}.  \jt{Journal of Fluid Mechanics}  \bvol{287},  \pg{317–348}.

\bibitem[Head \& Bandyopadhyay(1981)]{Bandyopadhyay981}
{\sc \au{Head, M.~R.} \& \au{Bandyopadhyay, P.}} \yr{1981}  \at{New aspects of turbulent boundary-layer structure}.  \jt{J. Fluid Mech.}  \bvol{107},  \pg{297--338}.

\bibitem[Hunt \& Carruthers(1990)]{Hunt1990}
{\sc \au{Hunt, J. C.~R.} \& \au{Carruthers, D.~J.}} \yr{1990}  \at{Rapid distortion theory and the ‘problems’ of turbulence.}  \jt{J. Fluid Mech.}  \bvol{212},  \pg{497--532}.

\bibitem[Hussain \& Ramjee(1976)]{Hussain1976}
{\sc \au{Hussain, A. K. M.~F.} \& \au{Ramjee, V.}} \yr{1976}  \at{Effects of the axisymmetric contraction shape on incompressible turbulent flow}.  \jt{J. Fluids. Eng.}  \bvol{98},  \pg{58--68}.

\bibitem[Hutchins \& Marusic(2007)]{Marusic_2007}
{\sc \au{Hutchins, N.} \& \au{Marusic, I.}} \yr{2007}  \at{Evidence of very long meandering features in the logarithmic region of turbulent boundary layers}.  \jt{J. Fluid Mech.}  \bvol{579},  \pg{1--28}.

\bibitem[Ianiro {\em et~al.\/}(2018)Ianiro, Lynch, Violato, Cardone \& Scarano]{Ianiro2018}
{\sc \au{Ianiro, A.}, \au{Lynch, K.~P.}, \au{Violato, D.}, \au{Cardone, G.} \& \au{Scarano, F.}} \yr{2018}  \at{Three-dimensional organization and dynamics of vortices in multichannel swirling jets}.  \jt{J. Fluid Mech.}  \bvol{843},  \pg{180--210}.

\bibitem[Kim {\em et~al.\/}(1971)Kim, Kline \& Reynolds]{Kim_Kline_Reynolds_1971}
{\sc \au{Kim, H.~T.}, \au{Kline, S.~J.} \& \au{Reynolds, W.~C.}} \yr{1971}  \at{The production of turbulence near a smooth wall in a turbulent boundary layer}.  \jt{Journal of Fluid Mechanics}  \bvol{50}~(1),  \pg{133–160}.

\bibitem[Marusic \& Monty(2019)]{Marusic2019}
{\sc \au{Marusic, I.} \& \au{Monty, J.~P.}} \yr{2019}  \at{Attached eddy model of wall turbulence}.  \jt{Annu. Rev. Fluid Mech.}  \bvol{51},  \pg{49--74}.

\bibitem[McKeon(2017)]{McKeon_2017}
{\sc \au{McKeon, B.~J.}} \yr{2017}  \at{The engine behind (wall) turbulence: perspectives on scale interactions}.  \jt{Journal of Fluid Mechanics}  \bvol{817},  \pg{P1}.

\bibitem[Mugundhan {\em et~al.\/}(2020)Mugundhan, Pugazenthi, Speirs, Samtaney \& Thoroddsen]{mugundhan2020alignment}
{\sc \au{Mugundhan, V.}, \au{Pugazenthi, R.~S.}, \au{Speirs, N.~B.}, \au{Samtaney, R.} \& \au{Thoroddsen, S.~T.}} \yr{2020}  \at{The alignment of vortical structures in turbulent flow through a contraction}.  \jt{J. Fluid Mech.}  \bvol{884},  \pg{A5}.

\bibitem[Prandtl(1933)]{prandtl1933attaining}
{\sc \au{Prandtl, L.}} \yr{1933}  \at{Attaining a steady air stream in wind tunnels}.  \jt{NACA Techn. Memo.}  \bvol{726}.

\bibitem[Ribner \& Tucker(1952)]{ribner1952spectrum}
{\sc \au{Ribner, HS} \& \au{Tucker, M}} \yr{1952}  \at{Spectrum of turbulence in a contracting stream. naca rep. n 1113}.  \jt{NACA Report}  \bvol{1113},  \pg{99--115}.

\bibitem[Rowley {\em et~al.\/}(2009)Rowley, I., S., Schlatter \& Henningson]{Rowley2009}
{\sc \au{Rowley, C.~W.}, \au{I., Mezi\'c}, \au{S., Bagheri}, \au{Schlatter, P.} \& \au{Henningson, D.~S.}} \yr{2009}  \at{Spectral analysis of nonlinear flows}.  \jt{J. Fluid Mech.}  \bvol{641},  \pg{115--127}.

\bibitem[Schanz {\em et~al.\/}(2016)Schanz, Gesemann \& Schr{\"o}der]{schanz2016shake}
{\sc \au{Schanz, D.}, \au{Gesemann, S.} \& \au{Schr{\"o}der, A.}} \yr{2016}  \at{Shake-the-box: Lagrangian particle tracking at high particle image densities}.  \jt{Exp. Fluids}  \bvol{57}~(5),  \pg{1--27}.

\bibitem[Schmid(2010)]{Schmid2010}
{\sc \au{Schmid, P.~J.}} \yr{2010}  \at{Dynamic mode decomposition of numerical and experimental data}.  \jt{J. Fluid Mech.}  \bvol{656},  \pg{5--28}.

\bibitem[Schoppa \& Hussain(2002)]{SCHOPPA_HUSSAIN_2002}
{\sc \au{Schoppa, W.} \& \au{Hussain, F.}} \yr{2002}  \at{Coherent structure generation in near-wall turbulence}.  \jt{Journal of Fluid Mechanics}  \bvol{453},  \pg{57–108}.

\bibitem[Schr\"{o}der \& Schanz(2023)]{Schroder2023}
{\sc \au{Schr\"{o}der, A.} \& \au{Schanz, D.}} \yr{2023}  \at{3d lagrangian particle tracking in fluid mechanics.}  \jt{Annu. Rev. Fluid Mech.}  \bvol{55},  \pg{511--540}.

\bibitem[Sreenivasan \& Narasimha(1978)]{Sreenivasan1978}
{\sc \au{Sreenivasan, K.~R.} \& \au{Narasimha, R.}} \yr{1978}  \at{Rapid distortion of axisymmetric turbulence.}  \jt{J. Fluid Mech.}  \bvol{84},  \pg{497--516}.

\bibitem[Tan-Atichat {\em et~al.\/}(1980)Tan-Atichat, Nagib \& Drubka]{Tanatichat1980}
{\sc \au{Tan-Atichat, J.}, \au{Nagib, H.~M.} \& \au{Drubka, R.~E.}} \yr{1980}  \at{Effects of axisymmetric contractions on turbulence of various scales}.  \jt{NASA Tech. Rep.}  \pg{pp. 1--362}.

\bibitem[Taylor(1935)]{taylor1935turbulence}
{\sc \au{Taylor, GI}} \yr{1935}  \at{Turbulence in a contracting stream}.  \jt{J. Appl. Math. Mech./Z. angew. Math. Mech.}  \bvol{15},  \pg{91--96}.

\bibitem[Thoroddsen \& Van~Atta(1995)]{Thoroddsen1995}
{\sc \au{Thoroddsen, S.~T.} \& \au{Van~Atta, C.~W.}} \yr{1995}  \at{The effects of a vertical contraction on turbulence dynamics in a stably stratified fluid}.  \jt{J. Fluid Mech.}  \bvol{285},  \pg{371--406}.

\bibitem[Uberoi(1956)]{Uberoi1956}
{\sc \au{Uberoi, M.~S.}} \yr{1956}  \at{Effect of wind-tunnel contraction on free-stream turbulence}.  \jt{J. Aeronautical Sci.}  \bvol{23},  \pg{754--764}.

\bibitem[Wieneke(2008)]{Wieneke2008}
{\sc \au{Wieneke, B.}} \yr{2008}  \at{Volume self-calibration for 3-d particle image velocimetry}.  \jt{Exp. Fluid}  \bvol{45},  \pg{549--556}.

\bibitem[Zhou {\em et~al.\/}(1999)Zhou, Adrian, Balachandar \& Kendall]{Zhou_Adrian_1999}
{\sc \au{Zhou, J.}, \au{Adrian, R.~J.}, \au{Balachandar, S.} \& \au{Kendall, T.~M.}} \yr{1999}  \at{Mechanisms for generating coherent packets of hairpin vortices in channel flow}.  \jt{J. Fluid Mech.}  \bvol{387},  \pg{353}.

\end{thebibliography}
\bibliographystyle{jfm}

\end{document}


\renewcommand{\thefigure}{S\arabic{figure}}
\renewcommand{\thetable}{S\arabic{table}}
\renewcommand{\thepage}{S\arabic{page}}
\renewcommand{\thesection}{S\arabic{section}}
\setcounter{figure}{0}
\setcounter{table}{0}
\setcounter{page}{1}
\setcounter{section}{0}

\begin{center}
{\huge Supplementary Material}\vspace{0.25in}\\
{\Large for \enquote{Coherent turbulent structures in a rapid contraction}}\vspace{0.2in}\\
{\Large by Alhareth, Mugundhan, Langley \& Thoroddsen}\vspace{0.25in}\\
\end{center}

\section{Curve fitting of alignment PDFs}

\begin{figure}[htp!]
\centering
\begin{tabular}{ c c}
\includegraphics[width=0.45\textwidth]{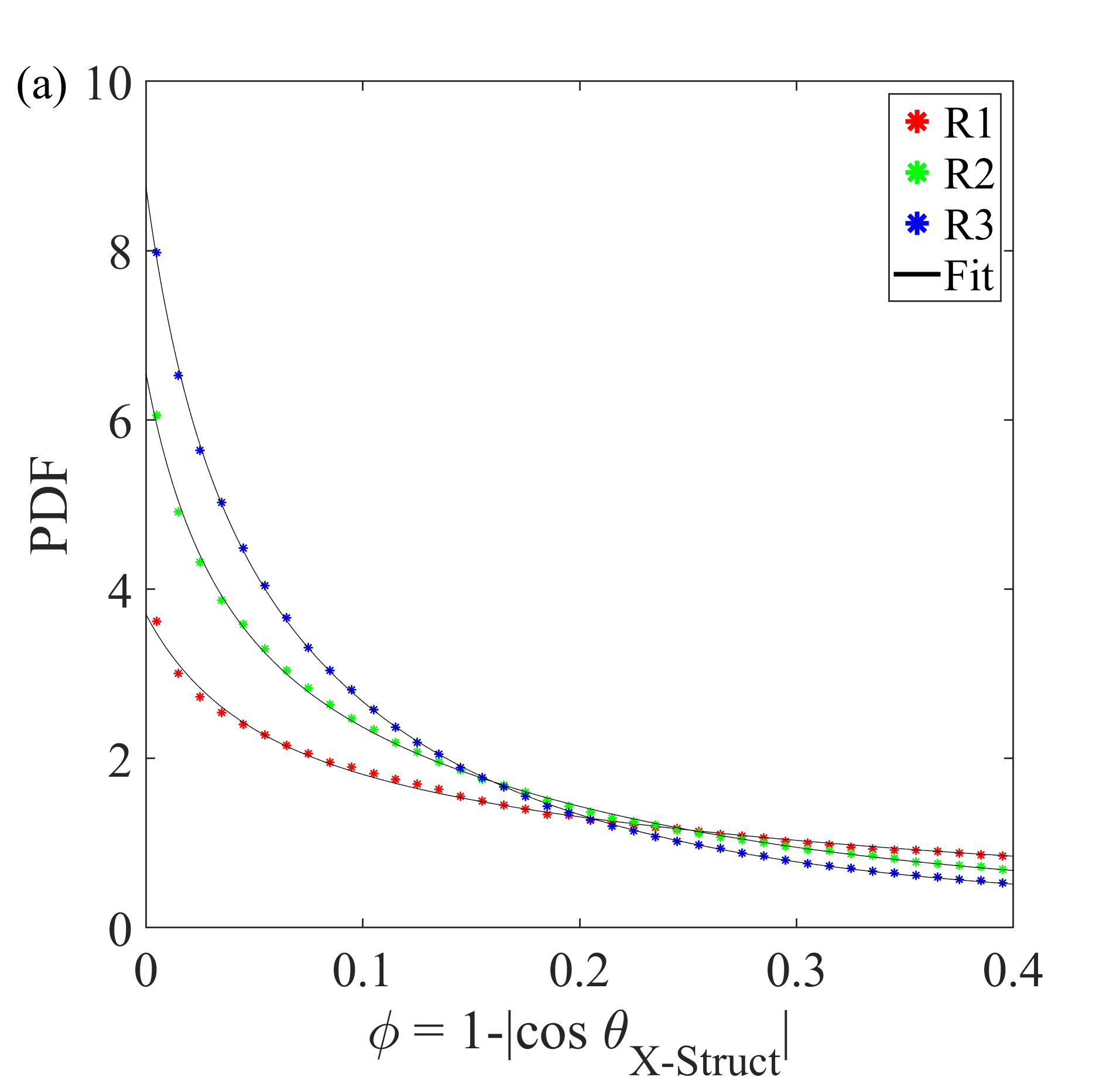} &
\includegraphics[width=0.45\textwidth]{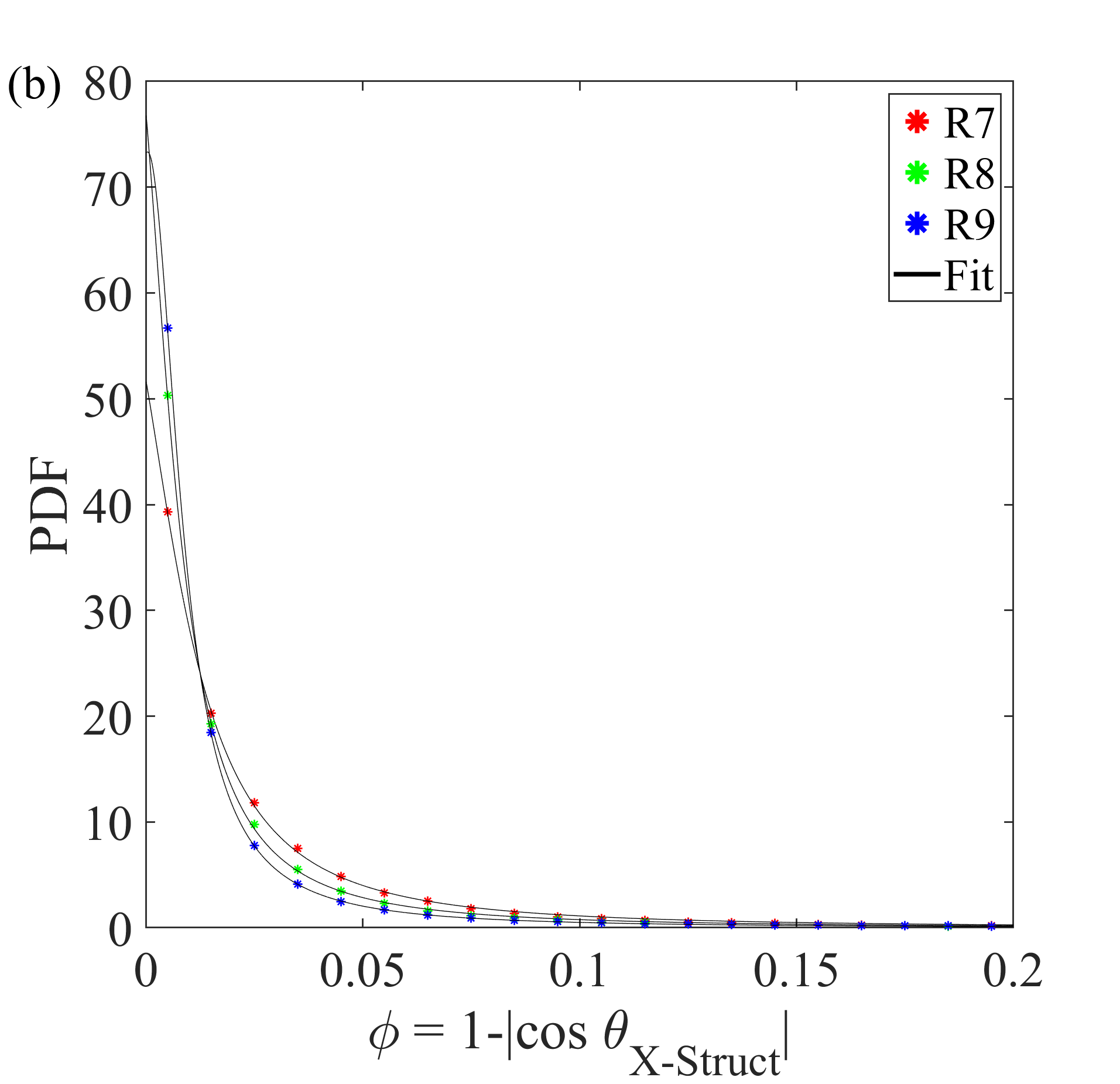}\vspace{-0.1in}\\
\label{Fit}
\end{tabular}
\caption{\textcolor{black}{The PDF and the inverted fifth-order polynomial fits obtained in volume sub-regions (a) R1-R3 inside region P1 near the inlet to the contraction; and (b) R7-R9 inside region P3 towards the exit of the contraction.
These PDFs are computed using 100 bins in the range [0,1]. 
 The cumulative probability of 99.5\% is reached before the fit reaches the last data point and thereby relies on the near-perfect interpolation of the fit.
 Thereby the computation of probability does not involve the value of fit at $\phi=$ 0, where the extrapolation is uncertain.}}
\end{figure}

\textcolor{black}{The alignment PDFs are strongly peaked around the streamwise direction.  To quantify the shape of the PDF of ${\left | cos \theta  \right |}$, we fit its peak near streamwise alignment with an inverted fifth-order polynomial in terms of $\phi = 1-{\left | cos \theta  \right |}$, as seen in Figure \ref{Fit}.  This fit works well for the full range [0,1].
The fits are of the form,}

\begin{equation}
PDF=\frac{1}{\phi^5+a_1\phi^4+a_2\phi^3+a_3\phi^2+a_4\phi++a_5}.
\end{equation}

\clearpage
\section{Volume of coherent vortical structure}

To ascertain the prevalence of the coherent vortical structures, we have calculated their volumes, using thresholding on the absolute value of the vorticity magnitude $\boldsymbol {\left | \omega  \right |}$= 40 s$^{-1}$

\begin{figure} [hbt!]
\centering
\includegraphics[width=0.95\textwidth]{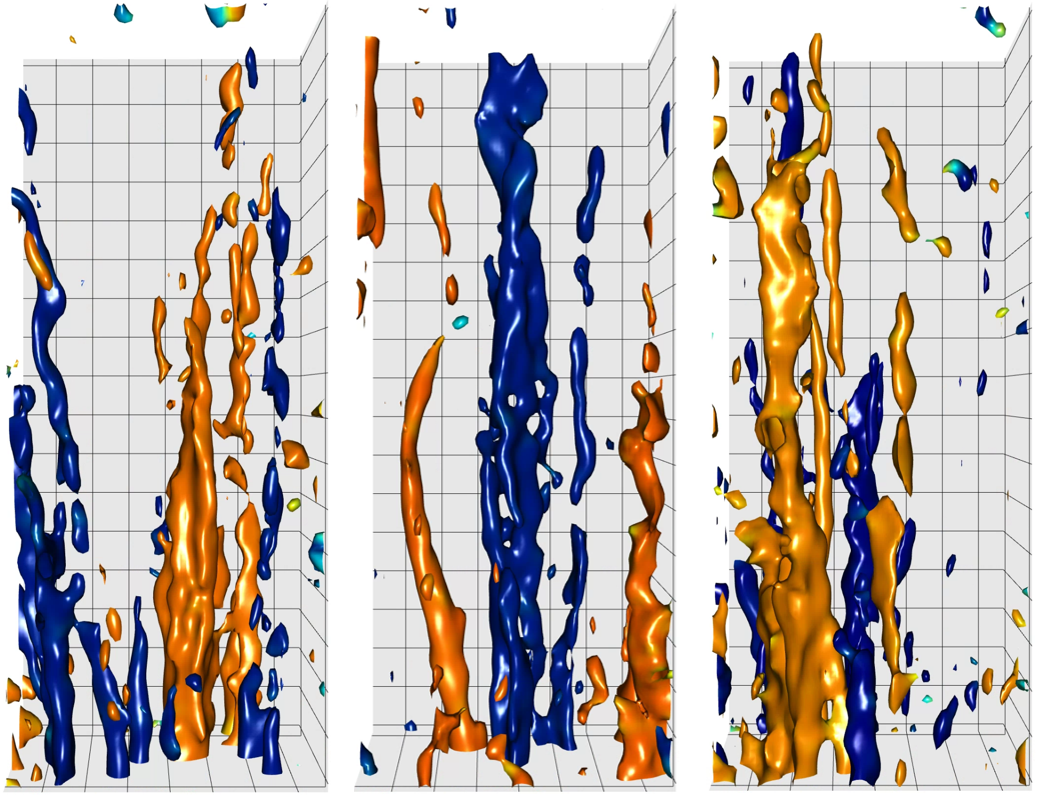}\vspace{-0.1in}\\
\caption{\textcolor{black}{The instantaneous coherent structures corresponding to the largest peak volume fractions marked by red dots in Figure 7 (a).  Using thresholding at $\boldsymbol {\left | \omega  \right |}$= 40 s$^{-1}$.}
}
\label{Volume_cs}
\end{figure}

\clearpage
\section{\textcolor{black}{Choice of threshold value for vorticity magnitude criterion}}
\label{sec:choice}

\begin{figure}[htp!]
\begin{center}
\begin{tabular}{ c}
\includegraphics[width=1.0\textwidth]{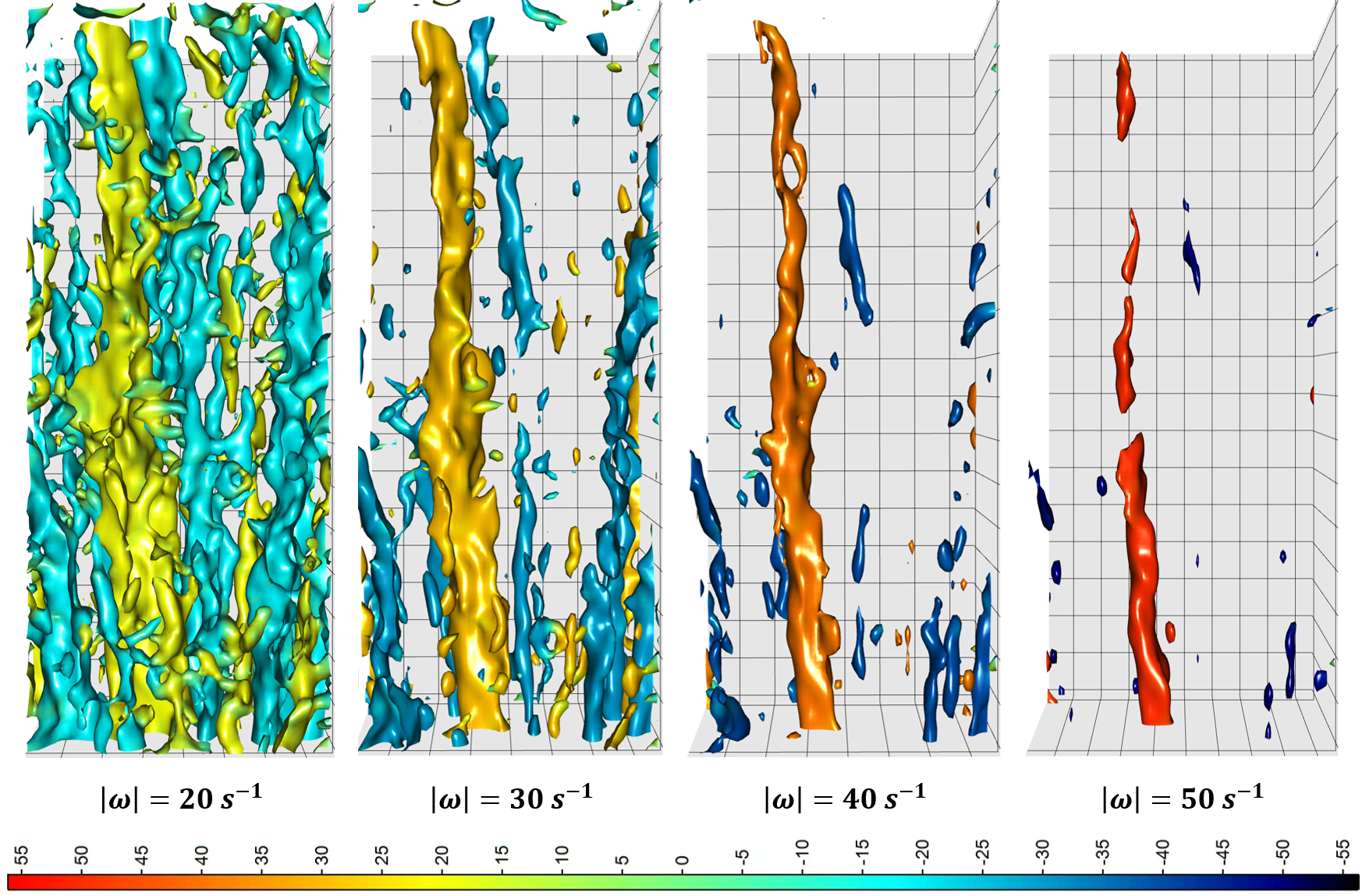}\vspace{-0.15in}\\
\end{tabular}
\end{center}
\caption{\textcolor{black}{Coherent vortical structures visualized with different isosurface thresholds corresponding to $|\omega|=$ 20$-$50 s$^{-1}$ ($|\omega|/S_{max}=$ 0.63$-$1.56) at the same instant in region P3. 
The isosurface is colored by $\omega_x$ whose color bar is shown.}}  
\label{CS_cutoff}
\end{figure}

\textcolor{black}{Figure \ref{CS_cutoff} shows how we select the threshold value for the prominent coherent streamwise vortical structure in region P3.  We compare the shapes for isosurfaces over a range of ${\left | \omega  \right |}=$ 20$-$50 s$^{-1}$, while only visualizing structures having a volume greater than 100 voxels.  For the lowest threshold value of 20 s$^{-1}$ the isosurfaces fill most of the volume, with irregular shapes.  The primary coherent structure is still visible, but random tendrils emerge from it and are not coherent in time.  This makes it difficult to identify distinct and dominant structures.  At 30 s$^{-1}$ the structure is more defined and regular, while numerous more discontinuous smaller structures persist.
On the other hand, for the highest threshold value of ${\left | \omega  \right |}=$50 s$^{-1}$ the vortex core breaks up, especially at the inlet, where it has not stretched as much as at the outlet.  This makes it hard to identify and track with time.
To isolate the coherent structures we thus arrive at an intermediate cut-off value of ${\left | \omega  \right |}=$40 s$^{-1}$.}

\section{\textcolor{black}{Fitting of coherent structures}}
\label{sec:fitting}

\begin{figure}[htp!]
\begin{center}
\begin{tabular}{ c}
(a) \\
\includegraphics[width=0.6\textwidth]{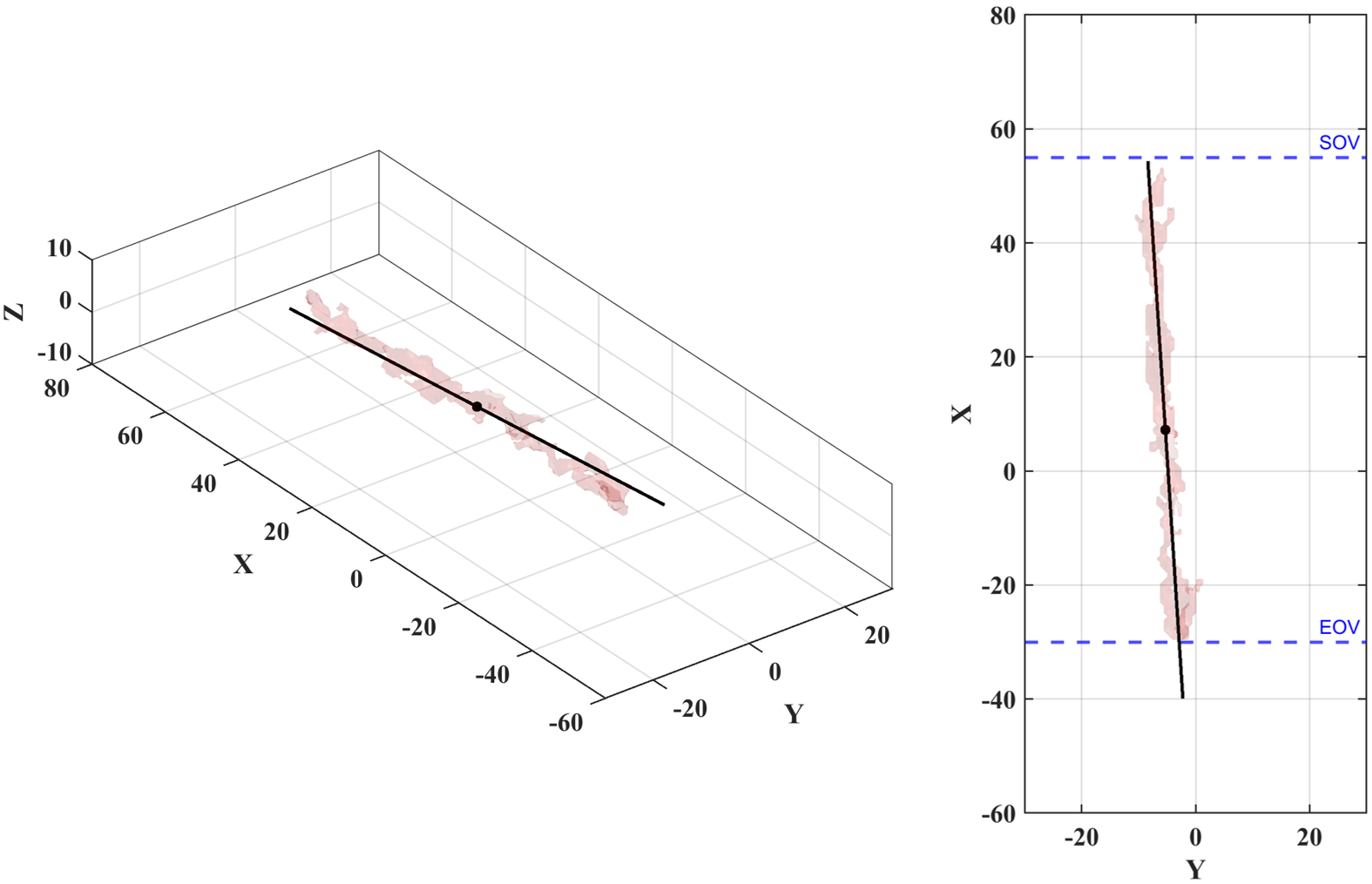}\\
(b) \\
\includegraphics[width=0.6\textwidth]{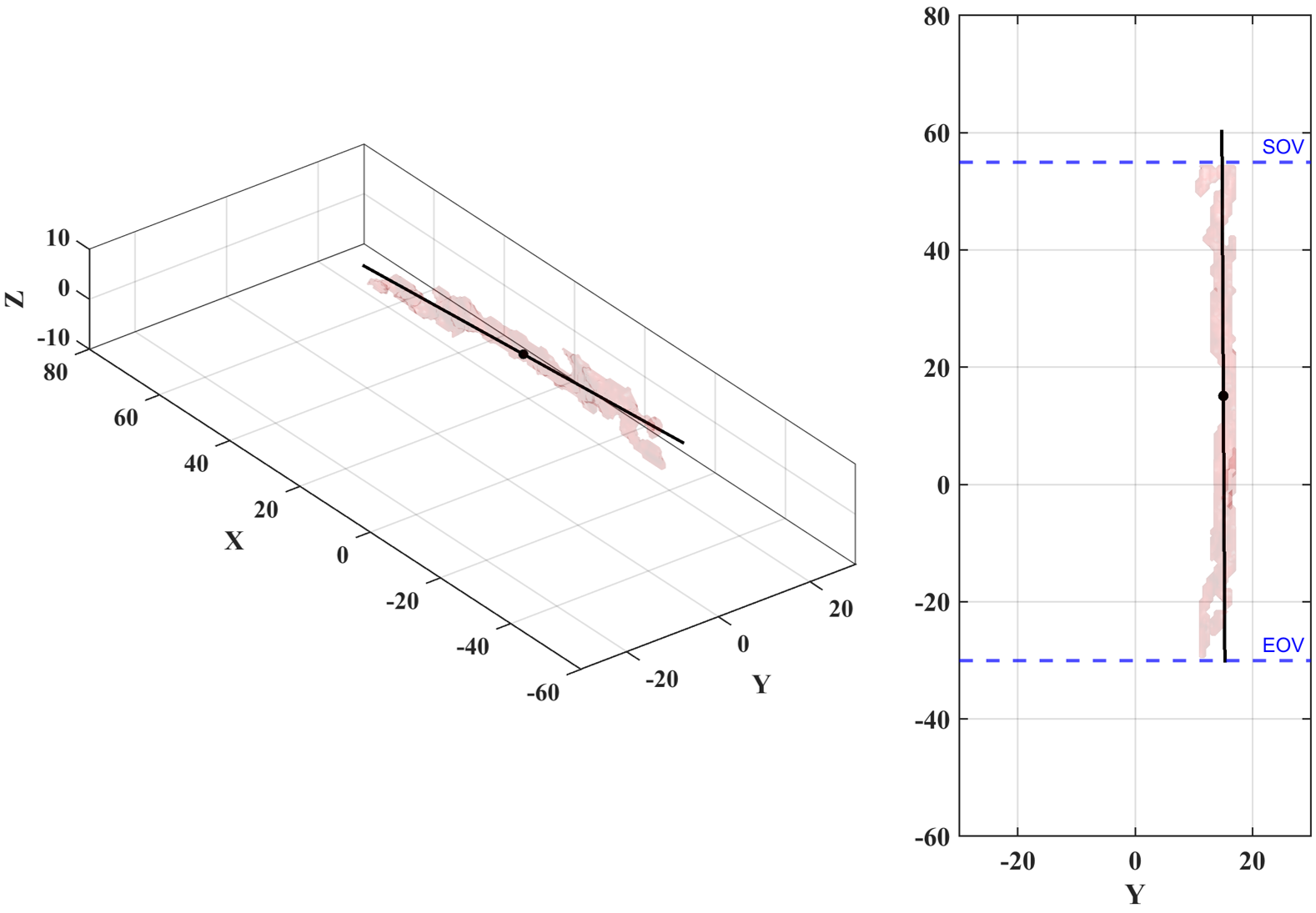}\\
(c) \\
\includegraphics[width=0.6\textwidth]{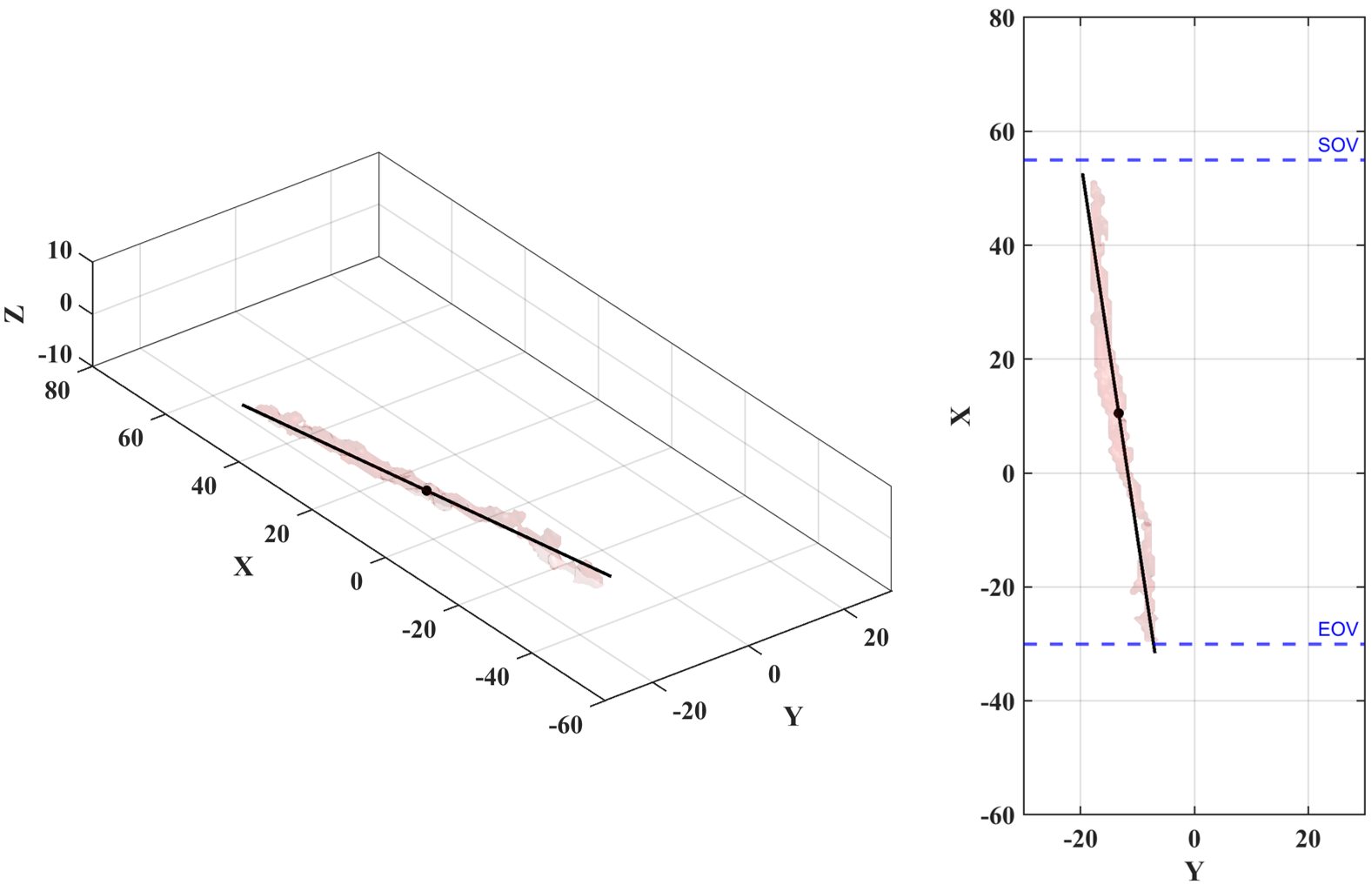}\\
\end{tabular}
\end{center}
\caption{\textcolor{black}{The 3-D views and the corresponding $XY$ projections for three independent, instantaneous coherent vortical structures (a$-$c) visualized within volume P3 using isosurface threshold of $|\omega|=$ 40 s$^{-1}$.
The centroid and the major principal axis of the equivalent ellipsoid fitted to structure is indicated by the black dot and the line.
The length of the line corresponds to the major-axis length $L$ of the ellipsoid.
$SOV$ and $EOV$ denote the start and end of the region P3.
The $XYZ$ coordinate system shown is local to region P3 as used in the spatial calibration.
The axis $X$ here corresponds to axis $-x$ shown in Figure 1 with a streamwise shift, whereas axes $Y,Z$ coincide with $y,z$.
The origin [0,0,0] in $XYZ$ coordinate corresponds to [296,0,0] mm in the $xyz$ coordinate system shown in Figure 1.}}
\label{CS_ellip}
\end{figure}

\textcolor{black}{We use the {\it watershed} algorithm in MATLAB to identify and tag the coherent structures visualized by isosurface of ${\left | \omega  \right |}$ and filtered based on the volume occupied by the structure.
The geometric properties of the tagged structures are extracted using the {\it regionprops3} function in MATLAB.
The function fits an equivalent ellipsoid which has the same normalized central moments as the structure. 
We use the major axis of this ellipsoid to characterize the alignment and length of the structures.
Figure \ref{CS_ellip} shows three examples of the longest, instantaneous structures identified in region P3, together with the centroid and major axis of the equivalent ellipsoid.}

\bibliographystyle{jfm}